\def\arxivbuild{1}  
\makeatletter
\def\input@path{{sections/}}
\makeatother

\documentclass[conference]{IEEEtran}
\IEEEoverridecommandlockouts

\usepackage{amsmath}
\usepackage{amsthm}
\usepackage{booktabs}
\usepackage{graphicx}
\usepackage{enumitem}
\setlist[itemize]{leftmargin=*, labelindent=0pt}
\setlist[enumerate]{leftmargin=*, labelindent=0pt}
\usepackage{xspace}
\usepackage{xcolor}

\usepackage{subcaption}
\usepackage{pgfplots}
\pgfplotsset{compat=1.17}
\usetikzlibrary{patterns}

\usepackage[ruled,vlined,linesnumbered]{algorithm2e}

\makeatletter
\newcommand{\heading}[1]{\par\addvspace{.5\baselineskip \@plus 2\p@ \@minus 2\p@}%
  \noindent\textbf{#1}\hspace{.5em}\ignorespaces}
\makeatother
\newcommand{\algfont}{\footnotesize}
\SetArgSty{textnormal}
\SetAlFnt{\algfont}
\SetAlCapFnt{\algfont}
\SetAlCapNameFnt{\algfont}

\definecolor{steelblue}{RGB}{70,130,180}
\definecolor{firebrick}{RGB}{178,34,34}
\definecolor{editcolor}{RGB}{0,0,255}

\newtheorem{definition}{Definition}

\newcommand{\sysname}{\textsf{Omakase}\xspace}

\ifdefined\submission
  \newcommand{\alberto}[1]{}
  \newcommand{\lef}[1]{}
  \newcommand{\ray}[1]{}
  \newcommand{\dakai}[1]{}
  \newcommand{\edited}[1]{#1}
\else
  \newcommand{\alberto}[1]{{\color{blue}\textbf{Alberto:} #1}}
  \newcommand{\lef}[1]{{\color{brown}\textbf{Lefteris:} #1}}
  \newcommand{\ray}[1]{{\color{cyan}\textbf{Ray:} #1}}
  \newcommand{\dakai}[1]{{\color{orange}\textbf{Dakai:} #1}}
  \newcommand{\edited}[1]{#1}  
\fi

\newif\ifarxiv
\ifdefined\arxivbuild\arxivtrue\else\arxivfalse\fi
\newcommand{\LR}[1]{\ifarxiv\ (Line~\ref{#1})\fi}        
\newcommand{\LRR}[2]{\ifarxiv\ (Lines~\ref{#1}--\ref{#2})\fi}
\newcommand{\LRc}[1]{\ifarxiv, Line~\ref{#1}\fi}         
\expandafter\def\csname fullalg@alg:conflict-graph\endcsname{1}
\expandafter\def\csname fullalg@alg:reorder-overview\endcsname{2}
\expandafter\def\csname fullalg@alg:simreorder\endcsname{3}
\expandafter\def\csname fullalg@alg:greedy-integration\endcsname{4}
\expandafter\def\csname fullalg@alg:merge-graphs-reorder\endcsname{5}
\expandafter\def\csname fullalg@alg:occ-dg\endcsname{6}
\newcommand{\FULLALG}[1]{Algorithm~\csname fullalg@#1\endcsname{} of the full version~\cite{omakase-full}}
\newcommand{\ALGP}[1]{\ifarxiv\ (Algorithm~\ref{#1})\else\ (\FULLALG{#1})\fi} 
\newcommand{\ALG}[2]{\ifarxiv Algorithm~\ref{#1}\else #2\fi}

\usepackage{tikz}
\usepackage{pgfplots}
\pgfplotsset{compat=1.17}

\pgfplotstableread{
theta Sequential OCC DepOcc IntDepOcc
0.01	4019.3	11212.08	11207.51	11195.06
0.5	4016.19	9810.98	9192.41	9230.36
0.99	4053.22	4767.14	5109.95	4816.64
}\ThetaThroughputTableA

\pgfplotstableread{
theta Sequential OCC DepOcc IntDepOcc
0.01	4031.17	11131.66	10292.44	9834.21
0.5	4023.77	7916.48	7777.01	10433.9
0.99	4036.75	4393.74	4630.92	5955.88
}\ThetaThroughputTableB

\pgfplotstableread{
theta Sequential OCC DepOcc IntDepOcc
0.01	4038.88	8934.75	8978.37	10839.88
0.5	4001.24	6822.94	6881.11	11284.23
0.99	4063.67	4090.18	4589.66	6946.06
}\ThetaThroughputTableC

\pgfplotstableread{
theta Sequential OCC DepOcc IntDepOcc
0.01	4038.88	8934.75	8978.37	10839.88
0.5	4001.24	6822.94	6881.11	11284.23
0.99	4063.67	4090.18	4589.66	6946.06
}\ThetaThroughputTableD

\pgfplotstableread{
theta Sequential OCC DepOcc IntDepOcc
0.01	1	2.789560371	2.788423357	2.785325803
0.5	1	2.442857534	2.288838426	2.298287681
0.99	1	1.176136504	1.260713704	1.188349016
}\ThetaSpeedUpTableA

\pgfplotstableread{
theta Sequential OCC DepOcc IntDepOcc
0.01	1	2.761396815	2.553214079	2.439542366
0.5	1	1.967428556	1.932767032	2.593065707
0.99	1	1.180312364	1.265189855	1.192568237
}\ThetaSpeedUpTableB

\pgfplotstableread{
theta Sequential OCC DepOcc IntDepOcc
0.01	1	2.212185061	2.222985085	2.683882661
0.5	1	1.967428556	1.932767032	2.593065707
0.99	1	1.00652366	1.129437184	1.709307104
}\ThetaSpeedUpTableC

\pgfplotstableread{
theta Sequential OCC DepOcc IntDepOcc
0.01	1	2.233919476	2.11343322	2.806902412
0.5	1	1.641902223	1.73890903	2.850015735
0.99	1	0.8670892337	1.012899588	2.186329306
}\ThetaSpeedUpTableD

\usepackage[hidelinks]{hyperref}

\begin{document}

\title{Omakase: Execution Congestion Control through Conflict-Graph Integration in Multi-Proposer Blockchain}

\author{%
  \IEEEauthorblockN{Dakai Kang\IEEEauthorrefmark{1},
    Ray Neiheiser\IEEEauthorrefmark{2},
    Alberto Sonnino\IEEEauthorrefmark{3}\IEEEauthorrefmark{4},
    Eleftherios Kokoris Kogias\IEEEauthorrefmark{3},
    Mohammad Sadoghi\IEEEauthorrefmark{5}}
  \IEEEauthorblockA{\IEEEauthorrefmark{1}Sei Labs\quad
    \IEEEauthorrefmark{2}Nanyang Technological University, Singapore\quad
    \IEEEauthorrefmark{3}Mysten Labs\quad
    \IEEEauthorrefmark{4}University College London (UCL)\\
    \IEEEauthorrefmark{5}Exploratory Systems Lab, University of California, Davis, USA}
  \IEEEauthorblockA{\small dakai@seinetwork.io, ray.neiheiser@ntu.edu.sg, \{alberto, lefteris\}@mystenlabs.com, msadoghi@ucdavis.edu}%
}

\maketitle

\begin{abstract}
  As multi-proposer protocols push consensus throughput, transaction execution has become the dominant bottleneck in blockchains. Layer-1 blockchains parallelize execution either \emph{pessimistically} (Solana, Sui), requiring upfront state-access declarations and serializing contended transactions, or \emph{optimistically} (Aptos, Sei, Monad), resolving dependencies at run time but suffering cascading aborts under contention. Both treat blocks in isolation, leaving inter-block parallelism unexploited and most cores idle.

  We introduce \sysname, a parallel execution framework that combines pre-execution, inter-block conflict-graph integration, and graph-aware Optimistic Concurrency Control (OCC). Replicas use idle cores to pre-execute only the blocks they propose, \edited{spreading proposer-side prediction work across concurrent proposers}. After building per-block conflict graphs and identifying their hot resources, a greedy scheduler integrates graphs across the blocks committed in a round, keeping an integration only when a lightweight balance heuristic predicts a parallelism gain. A graph-aware OCC executor then runs the integrated graph, admitting transactions in dependency order, which eliminates aborts when the predicted graph is exact and reduces re-executions when it is not.

  On 15{,}000 Ethereum mainnet blocks ($\sim$2.4M transactions) and synthetic workloads with configurable contention, \sysname reaches a 2.47$\times$ speedup over sequential execution, well above OCC and graph-aware baselines without inter-block integration. Against the strongest alternative, which runs the whole round as one optimistic window, \sysname matches mean throughput but is markedly more stable under high contention, because integration bounds how much optimistic work is in flight: this is the congestion control of our title. \edited{Integration is not free, but overlapping it with execution across rounds hides 94\% of the measured integration cost in the validator pipeline.}
\end{abstract}


\section{Introduction}

Modern blockchains have made dramatic progress in scaling consensus
throughput. \emph{Multi-proposer} protocols enable concurrent block proposals
and commit batches of blocks per round. This architecture spans DAG-based
protocols~\cite{dagrider,narwhal,bullshark,shoal,shoalpp,sailfish,duan2024dashing,mahimahi,fides}
(e.g., Mysticeti~\cite{mysticeti}, Bullshark~\cite{bullshark}), concurrent
consensus instances~\cite{rcc,spotless,mirbft,lyu2025orthrus,lyu2025ladon}
(e.g., RCC~\cite{rcc}), and multi-proposer dissemination
layers~\cite{autobahn,narwhal,ranchal2026hermes}. However, once consensus delivers a totally
ordered batch of blocks, replicas must still execute the transactions against
a shared state. As consensus throughput increases and latency drops, transaction execution has emerged
as the dominant bottleneck in end-to-end performance.

To address this bottleneck, high-performance Layer-1 blockchains adopt one of two parallel
execution architectures. The \emph{pessimistic} approach forces transactions to
declare state accesses upfront, allowing the scheduler to safely partition work
across cores. Solana's Sealevel~\cite{solana,sealevel} requires account-level
declarations, while Sui's Move runtime~\cite{mysten,sui-parallelization} relies
on object-level granularity. Conversely, the \emph{optimistic} approach
speculates and validates. Systems like Aptos~\cite{aptos}, Sei~\cite{sei}, and
Monad~\cite{monad} deploy Block-STM~\cite{blockstm}, an \emph{Optimistic Concurrency Control (OCC)} engine. Block-STM executes transactions concurrently against a multi-versioned state, relying on a sequential validator to abort and
re-execute transactions whose read sets are invalidated by earlier writers.

Both models present fundamental trade-offs. The pessimistic approach requires
runtimes tailored for explicit state declarations. This renders it incompatible
with the Ethereum Virtual Machine (EVM), where storage accesses depend on
runtime values and cannot be declared in advance. Furthermore, pessimistic
execution imposes a restrictive programming model; Sui, for instance, had to add \emph{dynamic objects} discovered at
runtime, which the scheduler can only execute sequentially. The optimistic
approach is more developer-friendly by avoiding upfront declarations.
However, while OCC performs better under low contention, its throughput degrades sharply
under heavy contention due to frequent aborts and re-executions.

\begin{figure*}[t]
  \centering
  \scalebox{0.9}{\begin{tikzpicture}[
      txnode/.style={draw=black!70, rounded corners=1pt, inner sep=0pt,
          minimum height=18pt, minimum width=18pt, font=\scriptsize},
      gantt/.style={draw=black!60, rounded corners=0.5pt,
          minimum height=14pt, minimum width=14pt,
          inner sep=0pt, font=\tiny},
      ptitle/.style={font=\small\bfseries, anchor=south},
      cpu/.style={font=\scriptsize, anchor=east},
      caplabel/.style={font=\scriptsize, anchor=north, text=gray!50!black},
    ]

    \node[ptitle] at (-4, -2) {$B_1$'s conflict graph};
    \node[txnode, fill=orange!50] (T1) at (-4, -2.5) {$T_1$};
    \node[txnode, fill=orange!50] (T2) at (-4, -4.1) {$T_2$};
    \node[txnode, fill=blue!50]   (T3) at (-2.8, -3.3) {$T_3$};
    \draw[->, semithick] (T1) -- (T2);

    \node[ptitle] at (-4, -5.2) {$B_2$'s conflict graph};
    \node[txnode, fill=green!50] (T4) at (-4, -5.6) {$T_4$};
    \node[txnode, fill=green!50] (T5) at (-4, -7.3) {$T_5$};
    \path[fill=blue!50]
    ([xshift=-9pt,yshift=-9pt]-2.8,-6.5) rectangle ([yshift=9pt]-2.8,-6.5);
    \path[fill=green!50]
    ([yshift=-9pt]-2.8,-6.5) rectangle ([xshift=9pt,yshift=9pt]-2.8,-6.5);
    \node[draw=black!70, rounded corners=1pt, inner sep=0pt,
      minimum height=18pt, minimum width=18pt,
      font=\scriptsize] (T6) at (-2.8, -6.5) {$T_6$};
    \draw[->, semithick] (T4) -- (T5);
    \draw[->, semithick] (T4) -- (T6);
    \draw[->, semithick] (T5) -- (T6);
    \draw[->, semithick, dashed, red!75!black] (T3) to[bend left=35] (T6);
    \node[font=\scriptsize, color=red!75!black] at (-2, -4.4)
    {\itshape inter-block edge};
    \node[font=\scriptsize, color=red!75!black] at (-2, -4.7)
    {\itshape (added by \sysname)};

    \begin{scope}[yshift=-3.2cm, xshift=0cm]
      \node[ptitle] at (3.2, 1.2) {(a) Sequential ($1$ CPU)};
      \node[cpu] at (0.55, 0.5) {CPU\,1};
      \foreach \i/\fill in {1/{orange!50}, 2/{orange!50}, 3/{blue!50},
          4/{green!50}, 5/{green!50}} {
          \pgfmathsetmacro{\xc}{0.8 + 0.55*\i}
          \node[gantt, fill=\fill] at (\xc, 0.5) {$T_\i$};
        }
      \pgfmathsetmacro{\xs}{0.8 + 0.55*6}
      \path[fill=blue!50]
      ([xshift=-7pt,yshift=-7pt]\xs,0.5)
      rectangle ([yshift=7pt]\xs,0.5);
      \path[fill=green!50]
      ([yshift=-7pt]\xs,0.5)
      rectangle ([xshift=7pt,yshift=7pt]\xs,0.5);
      \draw[black!60, rounded corners=0.5pt]
      ([xshift=-7pt,yshift=-7pt]\xs,0.5) rectangle
        ([xshift=7pt,yshift=7pt]\xs,0.5);
      \node[font=\tiny] at (\xs, 0.5) {$T_6$};
      \draw[->, gray] (1.05, 0) -- (5.5, 0) node[right,font=\scriptsize] {$t$};
      \foreach \i in {0,1,...,6} {
          \pgfmathsetmacro{\xt}{1.05 + 0.55*\i}
          \draw[gray!50] (\xt, 0) -- (\xt, -0.06);
          \node[font=\tiny, gray] at (\xt, -0.18) {\i};
        }
      \draw[dotted, gray!70] (2.70, 0) -- (2.70, 1);   
      \draw[dashed, gray!70] (4.35, 0) -- (4.35, 1);   
      \node[caplabel] at (3.0, -0.45) {Total: 6 units (1$\times$)};
    \end{scope}

    \begin{scope}[yshift=-3.2cm, xshift=7cm]
      \node[ptitle] at (3.0, 1.2) {(b) Block-STM ($3$ CPUs, per-block)};
      \foreach \i/\y in {1/1.05, 2/0.55, 3/0.05} {
          \node[cpu] at (0.55, \y) {CPU\,\i};
        }
      \node[gantt, fill=orange!50] at (1.35, 1.05) {$T_1$};
      \node[gantt, fill=orange!50, pattern=north east lines,
        pattern color=red!70] at (1.35, 0.55) {$T_2$};
      \node[gantt, fill=blue!50] at (1.35, 0.05) {$T_3$};
      \node[gantt, fill=orange!50] at (1.9, 1.05) {$T_2$};
      \draw[dotted, gray!70] (2.15, -0.4) -- (2.15, 1.4);
      \node[gantt, fill=green!50] at (2.45, 1.05) {$T_4$};
      \node[gantt, fill=green!50, pattern=north east lines,
        pattern color=red!70] at (2.45, 0.55) {$T_5$};
      \path[fill=blue!50]
      ([xshift=-7pt,yshift=-7pt]2.45,0.05)
      rectangle ([yshift=7pt]2.45,0.05);
      \path[fill=green!50]
      ([yshift=-7pt]2.45,0.05)
      rectangle ([xshift=7pt,yshift=7pt]2.45,0.05);
      \path[pattern=north east lines, pattern color=red!70]
      ([xshift=-7pt,yshift=-7pt]2.45,0.05) rectangle
        ([xshift=7pt,yshift=7pt]2.45,0.05);
      \draw[black!60, rounded corners=0.5pt]
      ([xshift=-7pt,yshift=-7pt]2.45,0.05) rectangle
        ([xshift=7pt,yshift=7pt]2.45,0.05);
      \node[font=\tiny] at (2.45, 0.05) {$T_6$};
      \node[gantt, fill=green!50] at (3.0, 1.05) {$T_5$};
      \path[fill=blue!50]
      ([xshift=-7pt,yshift=-7pt]3.0,0.55)
      rectangle ([yshift=7pt]3.0,0.55);
      \path[fill=green!50]
      ([yshift=-7pt]3.0,0.55)
      rectangle ([xshift=7pt,yshift=7pt]3.0,0.55);
      \path[pattern=north east lines, pattern color=red!70]
      ([xshift=-7pt,yshift=-7pt]3.0,0.55) rectangle
        ([xshift=7pt,yshift=7pt]3.0,0.55);
      \draw[black!60, rounded corners=0.5pt]
      ([xshift=-7pt,yshift=-7pt]3.0,0.55) rectangle
        ([xshift=7pt,yshift=7pt]3.0,0.55);
      \node[font=\tiny] at (3.0, 0.55) {$T_6$};
      \path[fill=blue!50]
      ([xshift=-7pt,yshift=-7pt]3.55,1.05)
      rectangle ([yshift=7pt]3.55,1.05);
      \path[fill=green!50]
      ([yshift=-7pt]3.55,1.05)
      rectangle ([xshift=7pt,yshift=7pt]3.55,1.05);
      \draw[black!60, rounded corners=0.5pt]
      ([xshift=-7pt,yshift=-7pt]3.55,1.05) rectangle
        ([xshift=7pt,yshift=7pt]3.55,1.05);
      \node[font=\tiny] at (3.55, 1.05) {$T_6$};
      \draw[->, gray] (1.05, -0.4) -- (5.5, -0.4) node[right,font=\scriptsize] {$t$};
      \foreach \i in {0,1,...,6} {
          \pgfmathsetmacro{\xt}{1.05 + 0.55*\i}
          \draw[gray!50] (\xt, -0.4) -- (\xt, -0.46);
          \node[font=\tiny, gray] at (\xt, -0.58) {\i};
        }
      \draw[dashed, gray!70] (3.8, -0.4) -- (3.8, 1.4);
      \node[caplabel] at (3.0, -0.85) {Total: 5 units (1.2$\times$), 4 wasted aborts};
    \end{scope}

    \begin{scope}[yshift=-6.5cm, xshift=0cm]
      \node[ptitle] at (3.2, 1.2) {(c) Graph-aware OCC ($3$ CPUs, per-block)};
      \foreach \i/\y in {1/1.05, 2/0.55, 3/0.05} {
          \node[cpu] at (0.55, \y) {CPU\,\i};
        }
      \node[gantt, fill=orange!50] at (1.35, 1.05) {$T_1$};
      \node[gantt, fill=orange!50] at (1.9,  1.05) {$T_2$};
      \node[gantt, fill=blue!50]   at (1.35, 0.55) {$T_3$};
      \draw[dotted, gray!70] (2.2, -0.4) -- (2.2, 1.4);
      \node[gantt, fill=green!50] at (2.45, 1.05) {$T_4$};
      \node[gantt, fill=green!50] at (3.0,  1.05) {$T_5$};
      \path[fill=blue!50]
      ([xshift=-7pt,yshift=-7pt]3.55,1.05)
      rectangle ([yshift=7pt]3.55,1.05);
      \path[fill=green!50]
      ([yshift=-7pt]3.55,1.05)
      rectangle ([xshift=7pt,yshift=7pt]3.55,1.05);
      \draw[black!60, rounded corners=0.5pt]
      ([xshift=-7pt,yshift=-7pt]3.55,1.05) rectangle
        ([xshift=7pt,yshift=7pt]3.55,1.05);
      \node[font=\tiny] at (3.55, 1.05) {$T_6$};
      \draw[->, gray] (1.05, -0.4) -- (5.5, -0.4) node[right,font=\scriptsize] {$t$};
      \foreach \i in {0,1,...,6} {
          \pgfmathsetmacro{\xt}{1.05 + 0.55*\i}
          \draw[gray!50] (\xt, -0.4) -- (\xt, -0.46);
          \node[font=\tiny, gray] at (\xt, -0.58) {\i};
        }
      \draw[dashed, gray!70] (3.8, -0.4) -- (3.8, 1.4);
      \node[caplabel] at (3.0, -0.85) {Total: 5 units (1.2$\times$), $B_2$ waits for $B_1$ commit};
    \end{scope}

    \begin{scope}[yshift=-6.5cm, xshift=7cm]
      \node[ptitle] at (3.0, 1.2) {(d) \sysname ($3$ CPUs, $B_1$ and $B_2$ integrated)};
      \foreach \i/\y in {1/1.05, 2/0.55, 3/0.05} {
          \node[cpu] at (0.55, \y) {CPU\,\i};
        }
      \node[gantt, fill=orange!50] at (1.35, 1.05) {$T_1$};
      \node[gantt, fill=blue!50]   at (1.35, 0.55) {$T_3$};
      \node[gantt, fill=green!50]  at (1.35, 0.05) {$T_4$};
      \node[gantt, fill=orange!50] at (1.9,  1.05) {$T_2$};
      \node[gantt, fill=green!50]  at (1.9,  0.05) {$T_5$};
      \path[fill=blue!50]
      ([xshift=-7pt,yshift=-7pt]2.45,1.05)
      rectangle ([yshift=7pt]2.45,1.05);
      \path[fill=green!50]
      ([yshift=-7pt]2.45,1.05)
      rectangle ([xshift=7pt,yshift=7pt]2.45,1.05);
      \draw[black!60, rounded corners=0.5pt]
      ([xshift=-7pt,yshift=-7pt]2.45,1.05) rectangle
        ([xshift=7pt,yshift=7pt]2.45,1.05);
      \node[font=\tiny] at (2.45, 1.05) {$T_6$};
      \draw[->, gray] (1.05, -0.4) -- (5.5, -0.4) node[right,font=\scriptsize] {$t$};
      \foreach \i in {0,1,...,4} {
          \pgfmathsetmacro{\xt}{1.05 + 0.55*\i}
          \draw[gray!50] (\xt, -0.4) -- (\xt, -0.46);
          \node[font=\tiny, gray] at (\xt, -0.58) {\i};
        }
      \draw[dashed, gray!70] (2.7, -0.4) -- (2.7, 1.4);
      \node[caplabel, text=teal!50!black] at (2.5, -0.85) {Total: 3 units (\textbf{2$\times$})};
    \end{scope}

  \end{tikzpicture}}

  \caption{A motivating example. Two blocks $B_1{=}\{T_1, T_2, T_3\}$ and
    $B_2{=}\{T_4, T_5, T_6\}$ are committed in one round; their per-block
    conflict graphs are colour-coded by dependency chain, with $T_6$
    split blue/green to mark its inter-block conflict with $T_3$. We
    show four execution modes on a $3$-CPU machine, with total time units
    labelled per panel.}
  \label{fig:motivating-example}
\end{figure*}
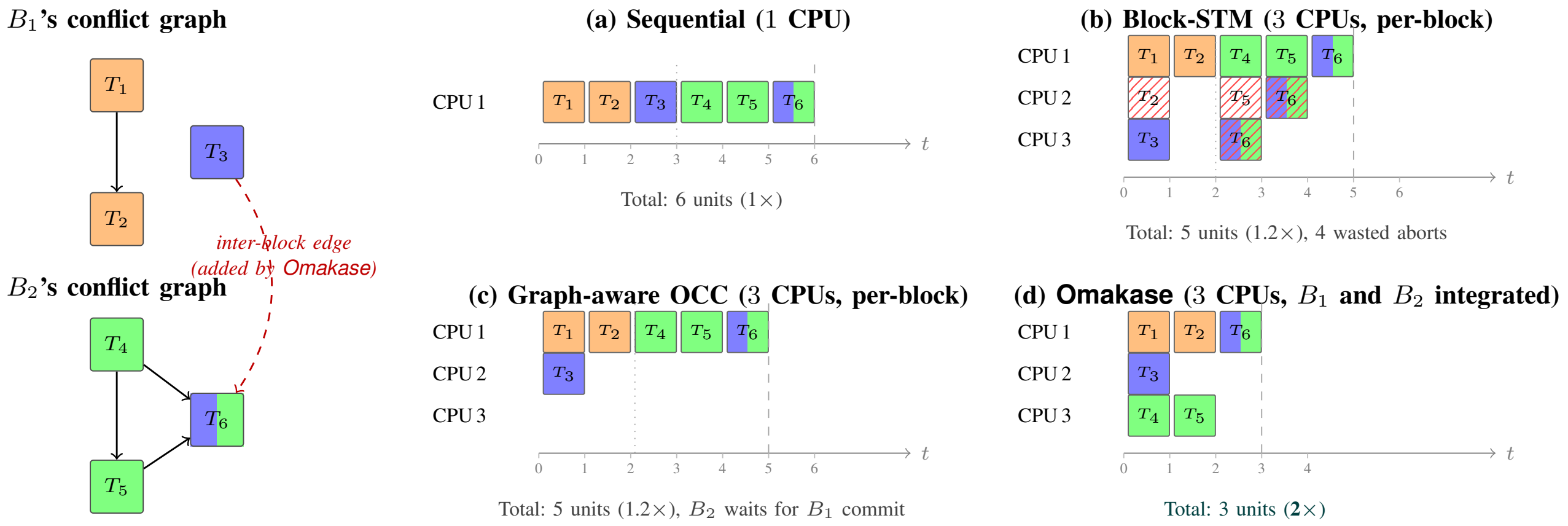

Moreover, although these engines already exploit \emph{intra-block}
parallelism, they still treat each block in isolation, leaving \emph{inter-block} parallelism untapped. 
Consequently, despite validator hardware
commonly having 16 or more cores, practical blockchain workloads limit
parallel speedup to only a few cores, leaving most CPU resources idle~\cite{chiron}.
Multi-proposer consensus protocols uniquely alter this landscape: because each
round commits a batch of blocks concurrently, they naturally expose both
intra-block and inter-block parallelism. Crucially, in a multi-proposer
setting, the computational cost of pre-executing a block can be cleanly
amortized across the network proposers. \S\ref{sec:overview} turns these observations into a design.

Building on these observations, we introduce \sysname, a parallel execution framework that
combines pre-execution, inter-block conflict-graph integration, and a
graph-aware OCC engine. Replicas leverage idle cores to pre-execute only the
blocks they propose: before broadcasting a block, the proposer executes it
to predict state accesses and construct a conflict graph.
A gas-weighted simulation then reorders the block's transactions so that
long dependency chains start early while every predicted conflict keeps
its block order, estimates the resulting per-thread load, and identifies hot
resources, whose access span dictates the critical path.
Upon committing a batch of blocks in a round, a greedy scheduler selectively
integrates their conflict graphs, retaining an integration only when a
lightweight heuristic predicts a parallelism gain, and re-simulates each
integrated graph to fix its execution order. Finally, rather than executing every transaction optimistically as in
Block-STM, our graph-aware OCC executor admits a transaction only after the
transactions it reads from, according to the conflict graph, have executed. A sequential validation phase then catches any dependency the graph missed. 

The title names what integration buys. Running a whole round as one
optimistic window is like a network without congestion control: fast
while conflicts are rare, collapsing when they are not, because every
conflict between blocks becomes a re-execution that can spread through
the window. \sysname bounds the window: \edited{it considers integrating a block to a group only when its hot-resource set is disjoint from that of the current integrated graph, and accepts the integration only if the balance test passes, so} the amount of optimistic work in flight is decided from the
graphs before execution, not discovered through aborts during it. The
price is some average throughput; the return is a round time that stays
close to the typical one even on a bad round.

We implement \sysname on top of Pevm~\cite{pevm} and evaluate it on
15{,}000 Ethereum mainnet blocks (about 2.4\,M transactions) and on
synthetic workloads with configurable contention
(\S\ref{sec:evaluation}). On real Ethereum \sysname executes a round
2.47$\times$ faster than sequential execution, against 1.89$\times$ for
Block-STM and 1.66$\times$ for the same graph-aware engine run block by
block. Running
the whole round as one optimistic window is about as fast on average
(2.40$\times$) but much less steady: its slowest round runs at
1.51$\times$ against 1.86$\times$ for \sysname. The integration step
costs more than it saves if a validator waits for it, but overlapping it
with the execution of the previous round hides 94\% of that cost, for a
validator-side speedup of 2.23$\times$ against 1.91$\times$ for
Block-STM.

\heading{Contributions.} This paper makes the following contributions:
\begin{itemize}
  \item A pre-execution + conflict-graph framework that exposes both
        intra-block and inter-block parallelism.
  \item A gas-weighted simulation that reorders a block's transactions
        along their dependency chains, estimates per-thread load, and
        identifies hot resources.
  \item A \emph{formulation} of inter-block graph integration as a
        balance-constrained scheduling problem, deciding which blocks' conflict graphs
        to integrate in a round so as to maximize thread utilization.
  \item \edited{A greedy heuristic for grouping conflict graphs using simulated load balance and hot-resource information.}
  \item A congestion-control view of optimistic execution across blocks:
        bounding the optimistic window keeps the round time stable where
        one large window collapses under inter-block conflict.
  \item An open-source implementation~\cite{code} on top of Pevm and an
        evaluation on real and synthetic Ethereum workloads against
        sequential execution, Block-STM, graph-aware OCC, Vegeta and the
        concatenated round, showing that \sysname outperforms every
        per-block engine and matches the concatenated round's throughput
        while remaining far more stable under contention, together with
        \edited{an accounting of what every stage costs and how overlapping integration with execution hides most of the integration cost in the validator pipeline.}
\end{itemize}


\section{Background and Related Work}
\label{sec:related}

Parallel execution is a widely explored topic in database research~\cite{paralleldb1,paralleldb2,paralleldb3,paralleldb4} and, more recently, in blockchains~\cite{hydra}. However, in the blockchain context the result of execution must be deterministic to guarantee that all validators converge on the same state, otherwise safety is violated. This narrows the relevant database literature to deterministic parallel execution systems such as Calvin and Bohm~\cite{serializabledb1,serializabledb2,serializabledb3,serializabledb4,quecc,caracal}. These engines, however, assume each transaction's read and write set is known before execution. This assumption does not hold for blockchain workloads such as EVM workloads: under the EVM, storage accesses depend on runtime state, so read/write sets cannot be fully known in advance and can at best be predicted.

Moreover, blockchain workloads are characterized by high contention and write-heavy access patterns~\cite{chiron}. This has motivated a body of blockchain-native work. BlockSTM~\cite{blockstm}, deployed in production by Aptos~\cite{aptos}, Sei~\cite{sei}, and Monad~\cite{monad}, is the most prominent example, executing transactions fully optimistically and resolving conflicts at runtime via a multi-version data store. Optimistic approaches, however, underperform under contention: because dependencies are unknown, transactions are executed optimistically and then validated in index order, so a transaction that read a value a lower-indexed transaction later overwrites is aborted and re-executed. Under heavy contention these aborts cascade, and at the limit throughput can fall below sequential execution.

To avoid the re-execution overhead that optimistic approaches incur under contention, a range of systems expose conflict information ahead of execution. One approach places this burden on the client: Solana~\cite{solana} and Sui~\cite{sui} mandate that each transaction declares all state it will read or write. These declarations are enforced as correctness conditions rather than treated as estimates, so a transaction that touches an undeclared location fails, while one that over-declares its footprint needlessly constrains
the scheduler.

A different approach shifts hint generation to the block proposer~\cite{polygonupdate,ethforward,blockpilotzhang2023,parallelsilva2023} where the leader executes transactions at block creation time to estimate the dependencies they will have during run-time. This is often done on the critical path of consensus~\cite{polygonupdate}, which results in deterministic hints but caps the throughput to the pre-execution speed.
To avoid this, a recent approach~\cite{anjana2025aft} removes the dependency on the proposer altogether by having each node independently infer conflict sets through static analysis of EVM and MoveVM bytecode before execution begins. Nevertheless, this analysis also needs to run on the critical path which significantly constrains the execution speed-up.

Multi-proposer consensus protocols~\cite{mysticeti,mirbft,lyu2025orthrus}, in which several nodes concurrently produce blocks each round, have grown in prominence, yet parallel execution has rarely been studied in this setting. Vegeta~\cite{vegeta} is the closest existing approach aimed at improving execution throughput in blockchains with multiple concurrent proposers. It proposes to spread the pre-execution workload across concurrent leaders to relieve the single-proposer bottleneck. Even so, multiple proposers in Vegeta serve only to pipeline hint generation for what ultimately becomes one block. Thunderbolt~\cite{thunderbolt} pre-executes transactions \emph{before} ordering: a transaction whose pre-executed read set fails validation causes its block to be discarded. We reimplement Vegeta and compare against it in \S\ref{sec:absolute}; Thunderbolt's block-discarding failure model has no counterpart in our setting, where every committed transaction must execute, so we position it analytically only. 

However, the parallelism that naturally arises from executing transactions across
independently proposed blocks simultaneously is still left unexploited. \sysname pursues precisely this opportunity: by treating concurrently proposed blocks as independent execution streams whose non-conflicting transactions can proceed in parallel, we maximize execution parallelism.

Work such as Anthemius~\cite{anthemius} that filter block content to maximize throughput is orthogonal and would further increase the cross-block parallelism potential that \sysname can leverage.
\section{System Model}
\label{sec:model}

We consider a multi-proposer blockchain with $N$ processes $p_1, \ldots, p_N$,
of which up to $f$ may be Byzantine~\cite{byzantinegeneral}. The consensus
protocol $\Pi$ is a black box: the proposers input blocks $B_1, \ldots, B_N$
and $\Pi$ outputs the same ordered sequence of blocks at every correct
process. Every correct process then executes the committed transactions,
which must be deterministic so that all processes reach the same state; as
in Ethereum, Aptos and Sui~\cite{ethereum,aptos,sui2022marlin}, execution
cost is measured in gas. Unlike prior work~\cite{anthemius, pilotfish,
remora}, we require neither clients nor the system to declare transaction
dependencies or gas upfront.

\heading{Transactions, resources, and gas.} A \emph{transaction} $T_i$ is a
deterministic program. Run against a state $\sigma$, it reads a set
$R_i(\sigma)$ of \emph{resources} (storage slots and account state), writes a
set $W_i(\sigma)$, and consumes gas $g_i(\sigma)$. The sets are
state-dependent and therefore cannot be known before execution; \sysname
predicts them by pre-executing the block and treats the resulting conflict
graph only as a scheduling hint (\S\ref{ss:pregraph}). Two transactions
\emph{conflict} on a resource when both access it and at least one writes it.

\heading{Concurrent execution.} An \emph{incarnation} $(i,c)$ is the $c$-th
execution attempt of $T_i$; a new incarnation starts whenever the previous
one fails validation. In a concurrent execution of a round $T_1,\dots,T_n$ on
$k$ threads, incarnations run in parallel against a \emph{multi-version}
state: an incarnation of $T_i$ reads each resource from the highest-indexed
writer below $i$ (or the pre-round state) and writes versions tagged
$(i,c)$. Validation then re-reads, in index order, every resource that
$T_i$'s latest incarnation read; if any value has since been overwritten by a
lower-indexed transaction, $T_i$ is re-executed (\S\ref{ss:occ}). The
execution \emph{commits} once every transaction has passed validation.

\heading{Serializability.} The committed state must equal that of executing
$T_1,\dots,T_n$ one by one in a single fixed order, which \sysname derives
deterministically from the round (\S\ref{ss:reorder}, \S\ref{ss:integrate}),
so every correct replica converges on the same state. The order keeps every
predicted conflict as the blocks have it (\S\ref{ss:reorder}). 

\section{System Overview}
\label{sec:overview}

\begin{figure}[t]
  \centering
  \scalebox{0.9}{\begin{tikzpicture}[
      proc/.style={draw=black!70, rounded corners=2pt, fill=blue!10,
          font=\scriptsize, align=center, inner sep=2pt, minimum height=0.6cm},
      mod3/.style={proc, fill=orange!20},
      mod4/.style={proc, fill=teal!20},
      hdr/.style={font=\scriptsize\bfseries\itshape, anchor=north west, inner sep=2pt},
      dlabel/.style={font=\tiny\itshape, text=gray!50!black, inner sep=1pt, align=left},
      arr/.style={->, semithick, gray!50!black},
    ]

    \draw[rounded corners=3pt, fill=blue!4, draw=black!50] (0,0) rectangle (8.6,2.45);
    \node[hdr] at (0,2.45)
    {Layer~1: Per-block stage at each proposer (before consensus)};
    \foreach \d in {0.2,0.1} {
      \draw[rounded corners=2pt, fill=white, draw=black!60]
      (0.3+\d,0.15+\d) rectangle (8.3+\d,1.8+\d);
    }
    \draw[rounded corners=2pt, fill=white, draw=black!70]
    (0.3,0.15) rectangle (8.3,1.8);
    \node[font=\scriptsize\bfseries, anchor=north west, inner sep=2pt] at (0.3,1.8)
    {Proposer $i$};
    \node[font=\tiny, text=gray!50!black, anchor=north east, inner sep=2pt] at (8.3,1.8)
    {$i = 1,\dots,N$};
    \node[proc, minimum width=1.9cm] (b1) at (1.7,0.85)
    {Pre-execute $B_i$:\\read/write sets,\\gas};
    \node[proc, minimum width=1.75cm] (b2) at (3.975,0.85)
    {Build conflict\\graph $G_i$:\\RAW/WAW/WAR};
    \node[proc, minimum width=2.5cm] (b3) at (6.55,0.85)
    {Simulate on $k$ threads:\\reorder, per-thread gas,\\hot resources};
    \draw[arr] (b1) -- (b2);
    \draw[arr] (b2) -- (b3);

    \draw[arr] (4.3,0.15) -- (4.3,-0.5);
    \node[dlabel, anchor=west] at (4.4,-0.17)
    {broadcast $B_i^*$: reordered txns, $G_i$,\\per-thread gas, hot resources};

    \draw[rounded corners=3pt, fill=gray!15, draw=black!50] (0,-0.5) rectangle (8.6,-0.95);
    \node[font=\scriptsize\bfseries] at (4.3,-0.725) {Layer~2: Multi-proposer consensus};

    \draw[arr] (4.3,-0.95) -- (4.3,-1.4);
    \node[dlabel, anchor=west] at (4.4,-1.175) {committed round of blocks};

    \draw[rounded corners=3pt, fill=orange!8, draw=black!50] (0,-1.4) rectangle (8.6,-3.3);
    \node[hdr] at (0,-1.4) {Layer~3: Per-round stage at every replica (after commit)};
    \node[mod3, minimum width=2.0cm] (gi1) at (1.75,-2.25)
    {Greedy grouping\\(hot-resource and\\balance tests)};
    \node[mod3, minimum width=2.0cm] (gi2) at (4.2,-2.25)
    {Integrate graphs:\\cross-block\\RAW/WAW/WAR edges};
    \node[mod3, minimum width=2.2cm] (gi3) at (6.75,-2.25)
    {Simulate integrated\\graph: reorder, CV,\\hot resources};
    \draw[arr] (gi1) -- (gi2);
    \draw[arr] (gi2) -- (gi3);
    \draw[arr] (gi3.south) -- ++(0,-0.22) -| (gi1.south);
    \node[dlabel, anchor=north] at (4.25,-2.97)
    {group's CV and hot resources for the next candidate};

    \draw[arr] (4.3,-3.3) -- (4.3,-3.75);
    \node[dlabel, anchor=west] at (4.4,-3.525) {integrated blocks with integrated graphs};

    \draw[rounded corners=3pt, fill=teal!8, draw=black!50] (0,-3.75) rectangle (8.6,-5.25);
    \node[hdr] at (0,-3.75) {Layer~4: Graph-aware OCC execution (per integrated block)};
    \node[mod4, minimum width=2.3cm] (l4a) at (2.3,-4.6)
    {Parallel execution:\\admit a txn once its\\RAW parents executed};
    \node[mod4, minimum width=2.7cm] (l4b) at (6.3,-4.6)
    {Validate in index order,\\re-execute on\\misprediction};
    \draw[arr] ([yshift=4pt]l4a.east) -- ([yshift=4pt]l4b.west);
    \node[dlabel, anchor=south] at (4.2,-4.45) {executed txns};
    \draw[arr] ([yshift=-4pt]l4b.west) -- ([yshift=-4pt]l4a.east);
    \node[dlabel, anchor=north] at (4.2,-4.75) {re-execute};
  \end{tikzpicture}}

  \caption{End-to-end architecture of \sysname.
    \textbf{Layer~1}: each of the $N$ proposers pre-executes the block it proposes, builds its conflict graph, simulates it on $k$ threads to
    reorder the transactions, estimate per-thread gas and identify hot
    resources, and broadcasts all of this with the block.
    \textbf{Layer~2}: consensus commits a round of blocks.
    \textbf{Layer~3}: every replica greedily integrates the committed
    graphs and re-simulates each integrated graph.
    \textbf{Layer~4}: each integrated block runs on the graph-aware OCC
    engine, whose sequential validation catches mispredictions.}
  \label{fig:overview-architecture}
\end{figure}
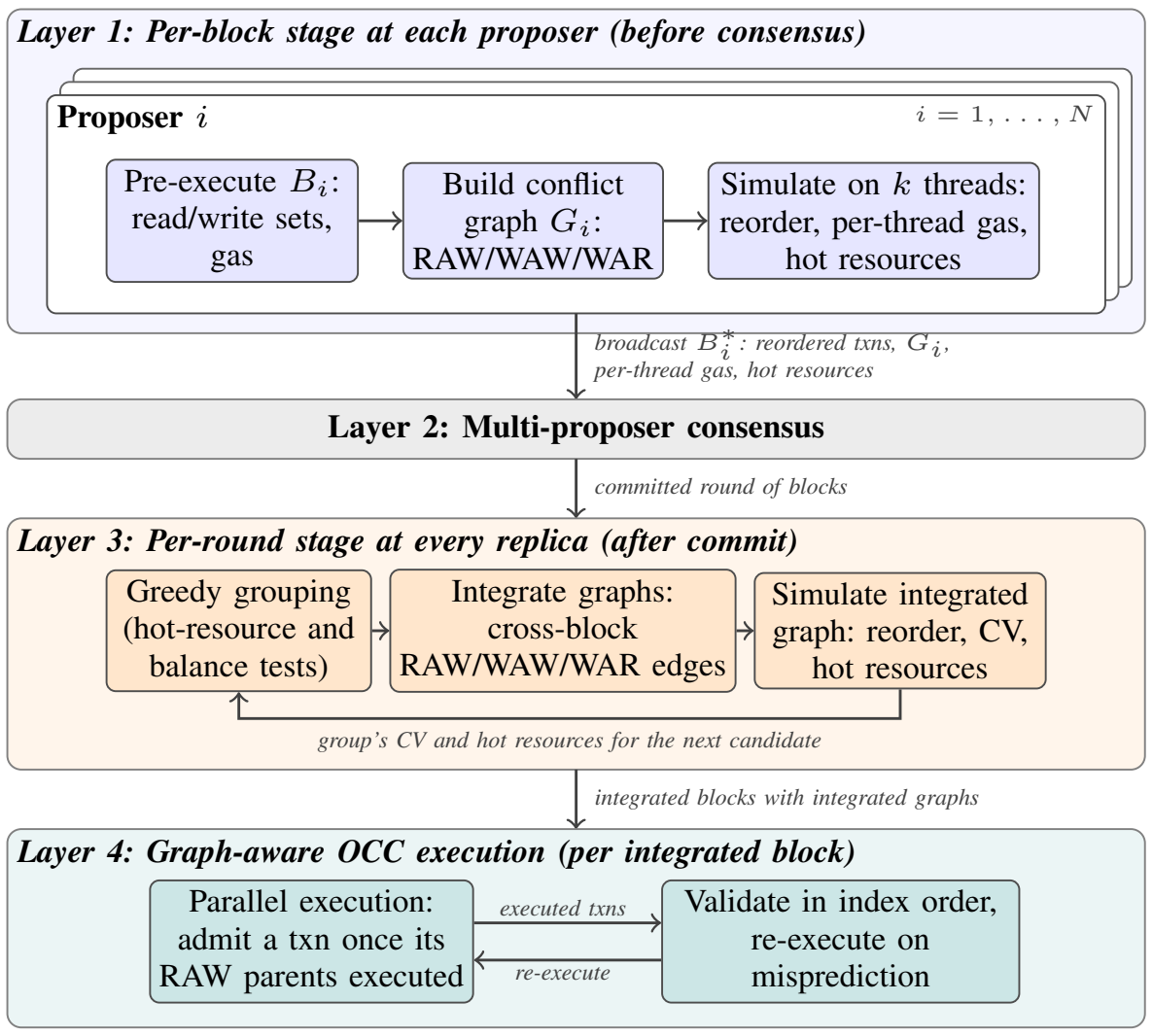

\sysname runs at every replica and slots into a multi-proposer consensus
protocol at two points: when a replica proposes a block, and when a round of
consensus commits a batch of blocks. It targets two opportunities that
current engines leave unused. \emph{Within} a block, OCC discovers conflicts
by aborting; a conflict graph predicted before execution lets the engine
schedule around most of them instead. \emph{Across} blocks, engines execute
one block at a time even when consecutive blocks touch disjoint hot
resources; the same predicted graphs make it possible to integrate such
blocks and run their transactions side by side.

Two observations make this practical.
\begin{enumerate}
  \item \emph{Prediction is affordable.} \edited{With balanced proposal rates and comparable block costs, each replica performs proposer-side prediction work for approximately $1/N$ of the committed blocks. This amortizes computational work across proposers.}
  \item \emph{The conflict graphs carry enough information to decide
        integration.} A block's graph, annotated with predicted gas,
        summarizes its dependencies: graphs of different blocks can be
        integrated without executing anything, their hot resources tell
        which pairs would only lengthen the critical path, and a
        single-threaded simulation that walks the integrated graph on $k$
        threads estimates per-thread load at a fraction of execution cost.
        \edited{Together, these summaries support a lightweight heuristic that selects integrations expected to improve worker utilization.}
\end{enumerate}
Figure~\ref{fig:overview-architecture} shows how the two stages realize them.

\heading{Per-block stage (before consensus).}
When a replica is about to broadcast a block, it pre-executes the block's
transactions against the latest committed state to predict each
transaction's read set, write set and gas. From these predictions it builds
the block's \emph{conflict graph}, whose nodes are transactions and whose
edges record every conflict on a shared resource: a read after a write,
two writes, or a write after a read~(\S\ref{ss:pregraph}). A
single-threaded scheduler then simulates parallel execution on $k$ threads
to reorder the transactions, keeping every predicted conflicting pair in
block order, to estimate per-thread load (\S\ref{ss:reorder}), and to identify \emph{hot resources}, whose
access span dominates~(\S\ref{ss:hot}). The graph, the reordered transaction list, the
per-thread load and the hot-resource set are broadcast with the block, so
the whole stage is paid once, by the proposer.

\heading{Per-round stage (after consensus commits).}
When a round commits a batch of blocks, every replica already holds each
block's conflict graph and hot-resource set. \edited{A greedy scheduler considers a block for integration to a group only when its hot-resource set is disjoint from that of the group's current integrated graph, and accepts the integration only if a balance test passes (\S\ref{ss:balance}, \S\ref{ss:integrate}).}
Each integrated graph is re-simulated to fix the execution order of its
transactions, exactly as a single block is (\S\ref{ss:reorder}), and then
executed by a graph-aware OCC engine that runs transactions in parallel
against a multi-version state and admits a transaction only after the
transactions it reads from have executed. Then a validation phase in
transaction-index order catches every misprediction (\S\ref{ss:occ}).
Groups are executed one after another, so a conflict between two
groups never goes through validation.

\heading{Why balanced integrations?}
Integration is constrained by the replica's hardware along two
dimensions. The single-core execution speed bounds the gas-weighted
\emph{depth} of a conflict graph: each chain of dependent transactions
must execute sequentially, though not necessarily on the same core, so its
gas is a lower bound on execution time. The core count $k$ bounds the
\emph{width}: how many chains can proceed side by side. \edited{An integration is therefore expected to pay off only when the integrated graphs} 
fill the cores that are idle otherwise, without making the critical path that a
single core must execute longer. This is the congestion control of our
title: the optimistic window should grow only when it fills idle cores.

\section{The \sysname Framework}

\subsection{Pre-Execution \& Conflict Graph Construction}\label{ss:pregraph}
\begin{figure*}[t]
  \centering
  \scalebox{0.95}{\begin{tikzpicture}[
      tx/.style={draw=black!70, rounded corners=1pt, inner sep=0pt,
          minimum height=17pt, minimum width=17pt, font=\scriptsize,
          fill=white},
      crit/.style={fill=orange!45},
      dep/.style={->, >=stealth, black!70, semithick},
      war/.style={->, >=stealth, black!70, semithick},
      elab/.style={font=\scriptsize, inner sep=1pt, fill=white},
      ptag/.style={font=\tiny, text=blue!55!black, inner sep=0.6pt,
          anchor=north east},
      ptitle/.style={font=\scriptsize\bfseries, anchor=south west},
      note/.style={font=\scriptsize, anchor=north west, inner sep=0pt},
      glab/.style={font=\tiny, anchor=west, inner sep=0pt},
      gbox/.style={draw=black!60, rounded corners=0.5pt, font=\scriptsize,
          inner sep=0pt},
    ]

    \begin{scope}
      \node[ptitle] at (0, 2.35) {(a) Alg.~1: pre-execution and conflict graph};
      \node[note] at (0, 2.1) {%
        \setlength{\tabcolsep}{2.2pt}\renewcommand{\arraystretch}{0.95}%
        \begin{tabular}{@{}lccc@{}}
          \toprule
          tx    & $R$ & $W$ & gas \\
          \midrule
          $T_1$ & $a$ & $a$ & 2 \\
          $T_2$ & $e$ & --  & 2 \\
          $T_3$ & --  & $b$ & 1 \\
          $T_4$ & $b$ & $c$ & 1 \\
          $T_5$ & $c$ & $d$ & 1 \\
          $T_6$ & $d$ & $e$ & 1 \\
          \bottomrule
        \end{tabular}};
      \node[tx] (aT1) at (2.4, 1.85) {$T_1$};
      \node[tx] (aT2) at (2.4, -0.7) {$T_2$};
      \node[tx] (aT3) at (3.3, 1.85) {$T_3$};
      \node[tx] (aT4) at (3.3, 1.0) {$T_4$};
      \node[tx] (aT5) at (3.3, 0.15) {$T_5$};
      \node[tx] (aT6) at (3.3, -0.7) {$T_6$};
      \draw[dep] (aT3) -- node[elab, right=1pt] {$b$} (aT4);
      \draw[dep] (aT4) -- node[elab, right=1pt] {$c$} (aT5);
      \draw[dep] (aT5) -- node[elab, right=1pt] {$d$} (aT6);
      \draw[war] (aT2) -- node[elab, above] {$e$} (aT6);
      \node[note] at (3.95, 2.1) {%
        \setlength{\tabcolsep}{1.6pt}\renewcommand{\arraystretch}{0.95}%
        \begin{tabular}{@{}llll@{}}
          \toprule
          $r$ & $\mathsf{Head}$ & $\mathsf{Tail}$ & $\mathsf{TailReaders}$ \\
          \midrule
          $a$ & $\{T_1\}$      & $T_1$ & $\emptyset$ \\
          $b$ & $\{T_3\}$      & $T_3$ & $\{T_4\}$ \\
          $c$ & $\{T_4\}$      & $T_4$ & $\{T_5\}$ \\
          $d$ & $\{T_5\}$      & $T_5$ & $\{T_6\}$ \\
          $e$ & $\{T_2, T_6\}$ & $T_6$ & $\emptyset$ \\
          \bottomrule
        \end{tabular}};
    \end{scope}

    \begin{scope}[xshift=7.9cm]
      \node[ptitle] at (0, 2.35) {(b) Alg.~2: $v.\mathsf{Path}$};
      \node[tx]       (bT1) at (0.3, 1.85) {$T_1$};
      \node[tx]       (bT2) at (0.3, -0.7) {$T_2$};
      \node[tx, crit] (bT3) at (1.3, 1.85) {$T_3$};
      \node[tx, crit] (bT4) at (1.3, 1.0) {$T_4$};
      \node[tx, crit] (bT5) at (1.3, 0.15) {$T_5$};
      \node[tx, crit] (bT6) at (1.3, -0.7) {$T_6$};
      \draw[dep, very thick, orange!80!black] (bT3) -- (bT4);
      \draw[dep, very thick, orange!80!black] (bT4) -- (bT5);
      \draw[dep, very thick, orange!80!black] (bT5) -- (bT6);
      \draw[war] (bT2) -- (bT6);
      \node[ptag] at (bT1.north east) {2};
      \node[ptag] at (bT2.north east) {3};
      \node[ptag] at (bT3.north east) {4};
      \node[ptag] at (bT4.north east) {3};
      \node[ptag] at (bT5.north east) {2};
      \node[ptag] at (bT6.north east) {1};
      \node[note] at (0, -1.15) {heap: $T_3{:}4,\ T_2{:}3,\ T_1{:}2$};
    \end{scope}

    \begin{scope}[xshift=10.2cm]
      \node[ptitle] at (0, 2.35) {(c) Alg.~3 vs.\ block order};
      \pgfmathsetmacro{\u}{0.40}   
      \pgfmathsetmacro{\xo}{0.9}   
      \pgfmathsetmacro{\h}{0.36}   
      \node[glab] at (0, 2.12) {no reordering, block order: 6 units};
      \node[glab] at (0.05, 1.77) {thr\,1};
      \node[glab] at (0.05, 1.35) {thr\,2};
      \draw[gbox, fill=white] (\xo, {1.77-\h/2}) rectangle ++(2*\u, \h) node[midway] {$T_1$};
      \draw[gbox, fill=orange!45] ({\xo+2*\u}, {1.77-\h/2}) rectangle ++(\u, \h) node[midway] {$T_3$};
      \draw[gbox, fill=orange!45] ({\xo+3*\u}, {1.77-\h/2}) rectangle ++(\u, \h) node[midway] {$T_4$};
      \draw[gbox, fill=orange!45] ({\xo+4*\u}, {1.77-\h/2}) rectangle ++(\u, \h) node[midway] {$T_5$};
      \draw[gbox, fill=orange!45] ({\xo+5*\u}, {1.77-\h/2}) rectangle ++(\u, \h) node[midway] {$T_6$};
      \draw[gbox, fill=white] (\xo, {1.35-\h/2}) rectangle ++(2*\u, \h) node[midway] {$T_2$};
      \node[glab, text=gray!50!black] at ({\xo+3.3*\u}, 1.35) {idle};
      \node[glab] at ({\xo+6*\u+0.1}, 1.77) {gas 6};
      \node[glab] at ({\xo+6*\u+0.1}, 1.35) {gas 2};
      \node[glab] at (0, 1.02) {reordered, longest path first: 4 units};
      \node[glab] at (0.05, 0.63) {thr\,1};
      \node[glab] at (0.05, 0.21) {thr\,2};
      \draw[gbox, fill=orange!45] (\xo, {0.63-\h/2}) rectangle ++(\u, \h) node[midway] {$T_3$};
      \draw[gbox, fill=orange!45] ({\xo+\u}, {0.63-\h/2}) rectangle ++(\u, \h) node[midway] {$T_4$};
      \draw[gbox, fill=orange!45] ({\xo+2*\u}, {0.63-\h/2}) rectangle ++(\u, \h) node[midway] {$T_5$};
      \draw[gbox, fill=orange!45] ({\xo+3*\u}, {0.63-\h/2}) rectangle ++(\u, \h) node[midway] {$T_6$};
      \draw[gbox, fill=white] (\xo, {0.21-\h/2}) rectangle ++(2*\u, \h) node[midway] {$T_2$};
      \draw[gbox, fill=white] ({\xo+2*\u}, {0.21-\h/2}) rectangle ++(2*\u, \h) node[midway] {$T_1$};
      \node[glab] at ({\xo+6*\u+0.1}, 0.63) {gas 4};
      \node[glab] at ({\xo+6*\u+0.1}, 0.21) {gas 4};
      \draw[->, gray] (\xo, -0.1) -- ({\xo+6.5*\u}, -0.1);
      \foreach \i in {0,...,6} {
        \draw[gray!60] ({\xo+\i*\u}, -0.1) -- ++(0, -0.05);
        \node[font=\tiny, gray!50!black] at ({\xo+\i*\u}, -0.25) {\i};
      }
      \node[font=\tiny, gray!50!black, anchor=west] at ({\xo+6.5*\u+0.05}, -0.1) {gas};
      \node[note, align=left] at (0, -0.45)
        {$\mathcal{L} = \langle T_3, T_4, T_2, T_5, T_6, T_1\rangle$\\
         $\mathsf{TotalGas} = (4, 4)$};
    \end{scope}

    \begin{scope}[xshift=14.3cm]
      \node[ptitle] at (0, 2.35) {(d) Defs.~1--2: spans, balance};
      \pgfmathsetmacro{\u}{0.36}   
      \pgfmathsetmacro{\xo}{0.45}  
      \foreach \r/\y/\s/\e in {a/1.95/2/4, b/1.65/0/2, c/1.35/1/3, d/1.05/2/4, e/0.75/0/4} {
        \node[font=\scriptsize, anchor=east] at ({\xo-0.08}, \y) {$\r$};
        \draw[fill=gray!35, draw=black!60] ({\xo+\s*\u}, {\y-0.1}) rectangle ({\xo+\e*\u}, {\y+0.1});
      }
      \draw[fill=red!45, draw=red!70!black] (\xo, 0.65) rectangle ({\xo+4*\u}, 0.85);
      \node[font=\tiny, anchor=west, text=red!70!black] at ({\xo+4*\u+0.05}, 0.75) {span 4: hot};
      \draw[red!70!black, densely dashed] ({\xo+3*\u}, 0.55) -- ({\xo+3*\u}, 2.12);
      \draw[->, gray] (\xo, 0.5) -- ({\xo+6.6*\u}, 0.5);
      \foreach \i in {0,2,4,6} {
        \draw[gray!60] ({\xo+\i*\u}, 0.5) -- ++(0, -0.05);
        \node[font=\tiny, gray!50!black] at ({\xo+\i*\u}, 0.35) {\i};
      }
      \node[font=\tiny, text=red!70!black, anchor=north west, inner sep=1pt] at (0, 0.22)
        {$\tau_{hot}\,\mathsf{MeanGas} = 0.75\cdot 4 = 3$};
      \node[note, text width=2.9cm, align=left] at (0, -0.15)
        {$e$ is hot ($4>3$), the others are not.
         $\mathsf{TotalGas}=(4,4)$: $\mathrm{CV}=0$, balanced.};
    \end{scope}
  \end{tikzpicture}}
  \caption{Algorithms 1--3 and the hot-resource test on a six-transaction
    block, with $k{=}2$ worker threads. (a)~Predicted accesses, the conflict graph
    and the per-resource maps. (b)~$v.\mathsf{Path}$, the longest path starting from each transaction node $v$, with
    the critical path highlighted. (c)~In block order the critical path starts at
    offset~2 and runs alone ($\mathsf{TotalGas}=(6,2)$); \ALG{alg:simreorder}{the simulation}
    starts it first ($\mathsf{TotalGas}=(4,4)$) and yields $\mathcal{L}$. (d)~Access spans in gas
    offsets: with $\tau_{hot}{=}0.75$, $e$ exceeds
    $\tau_{hot}\cdot\mathsf{MeanGas}=3$ and is hot. The $\mathsf{TotalGas}=(4,4)$
    gives this conflict graph $\mathrm{CV}=0$.}
  \label{fig:algos}
\end{figure*}
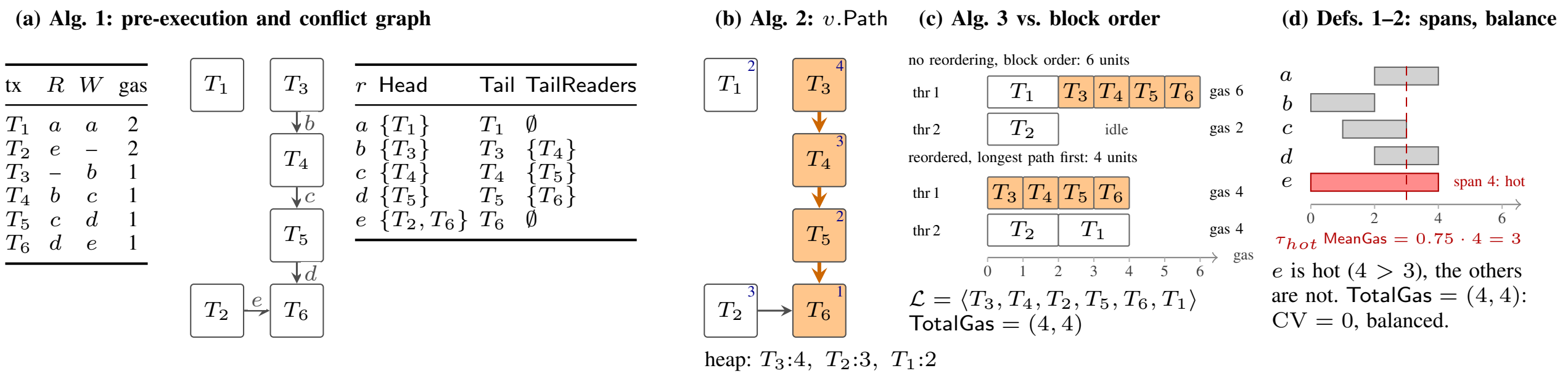

When proposing a block, a replica pre-executes its transactions to predict
each one's read set, write set and gas\ifarxiv\ (Algorithm~\ref{alg:conflict-graph},
Line~\ref{ln:cg-predict})\fi\footnote{We assume that no transaction fails due
to nonce mismatch.}, and builds the block's conflict graph by inserting the
transactions in the order they appear in the block\LR{ln:cg-fornode}.
Fig.~\ref{fig:algos} traces this and \ifarxiv the two algorithms\else the simulation\fi{} of
\S\ref{ss:reorder} on one block of 6 transactions.\ifarxiv\else{} The construction is
\FULLALG{alg:conflict-graph}, which also holds the pseudocode of every
other algorithm of this section.\fi

The graph records every conflict between two transactions of the block.
Each node is a transaction annotated with:
\begin{itemize}
  \item \textbf{Read set, write set, gas}: the predicted accesses and cost.
        The read set covers storage slots and account state (balance, nonce,
        code) alike, excluding only the block beneficiary.
  \item \textbf{Parents and children}: an edge $u \to v$ means that $u$
        precedes $v$ in the block and the two conflict on a resource: $v$
        reads what $u$ wrote (RAW), overwrites it (WAW), or writes what $u$
        read (WAR); $u$ is a parent of $v$ and $v$ a child of $u$. 
\end{itemize}
For each resource $r$ the construction maintains:
\begin{itemize}
  \item \textbf{Tail}: the latest writer of $r$. A new reader of $r$
        depends only on the tail, since an edge to the last writer
        subsumes all earlier accesses
       \LRR{ln:cg-reads-begin}{ln:cg-reads-end}; a new
        writer of $r$ depends on the tail (WAW\LRc{ln:cg-waw}) and
        becomes the tail\LRR{ln:cg-writes-begin}{ln:cg-writes-end}.
  \item \textbf{Tail readers}: the readers of $r$ since its last write. They
        are the WAR parents of $r$'s next writer, which clears the set
       \LR{ln:cg-war}.
  \item \textbf{Heads}: the transactions that access $r$ before its first
        write within the block, that is, $r$'s entry points with no incoming
        dependency on $r$. They are where another block's tail attaches to
        this graph when graphs are integrated (\S\ref{ss:integrate}).
\end{itemize}

The graph is a \emph{scheduling hint}, not the serialization mechanism. It
may be imprecise because a transaction's accesses depend on the state it
executes against: pre-execution runs the block against the state at
proposal time, whereas the committed execution runs it after the preceding
transactions of the round, including those of other blocks. If the state has changed,
the transaction may touch resources its prediction did not
include. Correctness never depends on the graph being exact (\S\ref{ss:occ}).

\ifarxiv\begin{algorithm}[t]
  \caption{Pre-execution and Conflict Graph Construction}
  \label{alg:conflict-graph}
  \DontPrintSemicolon
  \KwIn{Ordered block transactions $\mathcal{T}=\langle tx_1,\dots,tx_n\rangle$}
  \KwOut{Conflict graph $G=(V,E)$; per-resource maps $\mathsf{Head}$, $\mathsf{Tail}$, $\mathsf{TailReaders}$}

  $G\leftarrow (\emptyset,\emptyset)$\;
  $\mathsf{Head}\leftarrow \emptyset$ \tcp*{resource $\mapsto$ head nodes (default $\emptyset$)}
  $\mathsf{Tail}\leftarrow \emptyset$ \tcp*{resource $\mapsto$ last writer (default $\bot$)}
  $\mathsf{TailReaders}\leftarrow \emptyset$ \tcp*{resource $\mapsto$ readers since its last write}

  \BlankLine
  \For{$i\leftarrow 1$ \KwTo $n$}{\label{ln:cg-fornode}
    $(R_i,W_i,g_i)\leftarrow \textsc{SeqExecPredict}(tx_i)$\label{ln:cg-predict}\tcp*{seq.\ pre-exec of $tx_i$: predict $R,W$, gas}
    $v_i \leftarrow \textsc{AddNode}(G, tx_i)$;\ $v_i.R \leftarrow R_i$;\ $v_i.W \leftarrow W_i$;\ $v_i.gas \leftarrow g_i$\;
    $P\leftarrow \emptyset$ \tcp*{parents of $v_i$}

    \BlankLine
    \tcp{Reads: RAW edge from the last writer, else a new head}
    \ForEach{$r \in R_i$}{\label{ln:cg-reads-begin}
      \lIf{$\mathsf{Tail}[r]\neq \bot$}{$P\leftarrow P\cup\{\mathsf{Tail}[r]\}$}
      \lElse{$\mathsf{Head}[r]\leftarrow \mathsf{Head}[r]\cup\{v_i\}$}
      $\mathsf{TailReaders}[r]\leftarrow \mathsf{TailReaders}[r]\cup\{v_i\}$\;\label{ln:cg-reads-end}
    }

    \BlankLine
    \tcp{Writes: WAW edge from the last writer, WAR edges from its readers}
    \ForEach{$r \in W_i$}{\label{ln:cg-writes-begin}
      \lIf{$\mathsf{Tail}[r]\neq \bot$}{$P\leftarrow P\cup\{\mathsf{Tail}[r]\}$}\label{ln:cg-waw}
      \lElse{$\mathsf{Head}[r]\leftarrow \mathsf{Head}[r]\cup\{v_i\}$}
      $P\leftarrow P\cup \mathsf{TailReaders}[r]$\;\label{ln:cg-war}
      $\mathsf{Tail}[r]\leftarrow v_i$;\ $\mathsf{TailReaders}[r]\leftarrow \emptyset$\;\label{ln:cg-writes-end}
    }

    \BlankLine
    \ForEach{$u \in P\setminus\{v_i\}$}{\label{ln:cg-edges}
      $E \leftarrow E \cup \{(u,v_i)\}$;\ $u.\mathsf{Children}\leftarrow u.\mathsf{Children}\cup\{v_i\}$;\ $v_i.\mathsf{Parents}\leftarrow v_i.\mathsf{Parents}\cup\{u\}$\;
    }
  }

  \Return{$(G,\mathsf{Head},\mathsf{Tail},\mathsf{TailReaders})$}\;
\end{algorithm}
\fi

\subsection{Reordering Transactions via Simulation}\label{ss:reorder}

Consider the block of Fig.~\ref{fig:algos} on $k=2$ threads. Scheduled in
the order the transactions appear (Fig.~\ref{fig:algos}c, top), the two
independent transactions go first, and the chain
$T_3 \to T_4 \to T_5 \to T_6$ starts only at offset~2 and then runs alone
for four units while the other thread idles: the threads carry 6 and 2 gas
and the block takes 6 units. Starting the chain first (bottom) lets $T_1$
and $T_2$ fill the other thread: 4 and 4 gas, done at 4. \sysname reorders the block's transactions
by simulating their parallel execution on $k$ threads\ALGP{alg:reorder-overview}, scheduling the transaction with the
most expensive remaining chain first to shorten the critical path.

Concretely, it first computes for every node the cost of the most expensive
path starting there\LRR{ln:reorder-path-begin}{ln:reorder-path-end}: gas is
attached to nodes, so a path's cost is the sum of the gas of the
transactions on it, and a node's value is its own gas plus the maximum value
of its children, computed by depth-first search with memoization
(Fig.~\ref{fig:algos}b: $T_3$ has 4, and $T_2$ has 3 through its WAR edge
to $T_6$).

It then runs the simulation (\ifarxiv Line~\ref{ln:reorder-sim},
Algorithm~\ref{alg:simreorder}; \else\FULLALG{alg:simreorder}; \fi
Fig.~\ref{fig:algos}c, bottom):
transactions whose parents have all been scheduled are pushed into a
max-heap keyed by path cost\LRR{ln:reorder-heap-begin}{ln:reorder-heap-end}, initially
$T_3$, $T_2$ and $T_1$. Then an idle
simulated thread takes the top of the heap and advances by that
transaction's gas\ifarxiv\ (Algorithm~\ref{alg:simreorder},
Lines~\ref{ln:sim-assign-begin}--\ref{ln:sim-assign-end})\fi, and a transaction
joins the reordered list $\mathcal{L}$ when its simulated thread finishes it\LRR{ln:sim-exec-begin}{ln:sim-exec-end}, giving
$\mathcal{L} = \langle T_3, T_4, T_2, T_5, T_6, T_1\rangle$ in
Fig.~\ref{fig:algos}.

\heading{What the new order preserves.} $\mathcal{L}$ becomes the order
in which the block's transactions are serialized (\S\ref{sec:model}). Since
every predicted conflict is an edge, $\mathcal{L}$, a topological order of
the graph, keeps each such pair as the block has it. Validation (\S\ref{ss:occ}) then guarantees that the
committed state is that of executing the transactions one by one in
$\mathcal{L}$ order. The graph does not change with the order, only the
transaction indices do; the schedule the simulation computed, with its
per-thread gas and gas offsets, therefore describes the reordered block and
feeds \S\ref{ss:hot} and \S\ref{ss:balance}.

\ifarxiv\begin{algorithm}[t]
  \caption{TxnReorder}
  \label{alg:reorder-overview}
  \DontPrintSemicolon
  \KwIn{Conflict graph $G=(V,E)$; number of threads $k$}
  \KwOut{Reordered list of transaction nodes $\mathcal{L}$}

  \tcp{(1) Compute longest gas-weighted path from each node (DFS + memoization)}
  \ForEach{$v\in V$}{\label{ln:reorder-path-begin}
    $v.\mathsf{RemParents}\leftarrow |\;v.\mathsf{Parents}\;|$\;
    $v.\mathsf{Path}\leftarrow \bot$\;
  }
  \ForEach{$v\in V$}{
    \If{$v.\mathsf{Path}=\bot$}{
      $\textsc{DFSPath}(v)$\;\label{ln:reorder-path-end}
    }
  }

  \BlankLine
  \tcp{(2) Initialize max-heap keyed by $v.\mathsf{Path}$ with all nodes without parents}
  $\mathcal{H}\leftarrow \textsc{MaxHeap}()$\;\label{ln:reorder-heap-begin}
  \ForEach{$v\in V$}{
    \If{$v.\mathsf{RemParents}=0$}{
      $\textsc{Push}(\mathcal{H}, v,\; key=v.\mathsf{Path})$\;\label{ln:reorder-heap-end}
    }
  }

  \BlankLine
  \tcp{(3) Simulate $k$-thread execution and output the reordered list}
  $(\mathcal{L}, \mathsf{TotalGas})\leftarrow \textsc{SimReorder}(G,\mathcal{H},k)$\;\label{ln:reorder-sim}

  \Return{$(\mathcal{L}, \mathsf{TotalGas})$}\;
\end{algorithm}

\BlankLine
\begin{algorithm}[t]
  \caption{\textsc{SimReorder}: Single-Thread Simulation of $k$ Parallel Threads}
  \label{alg:simreorder}
  \DontPrintSemicolon
  \KwIn{Graph $G=(V,E)$; max-heap $\mathcal{H}$; number of threads $k$}
  \KwOut{Reordered list $\mathcal{L}$; per-thread total execution time $\mathsf{TotalGas}$}

  $\mathcal{L}\leftarrow [\,]$\;
  \For{$t\leftarrow 1$ \KwTo $k$}{
    $\mathsf{Offset}[t]\leftarrow 0$\tcp*{gas offset when $t$ is free}
    $\mathsf{Assigned}[t]\leftarrow \bot$\tcp*{assigned node}
    $\mathsf{TotalGas}[t]\leftarrow 0$\tcp*{total gas executed by thread $t$}
  }

  \While{$|\mathcal{L}| < |V|$}{
    \BlankLine
    \tcp{Assign ready nodes to idle threads (smallest idle thread id)}
    \While{$\exists t:\ \mathsf{Assigned}[t]=\bot \ \wedge\ \neg\textsc{Empty}(\mathcal{H})$}{\label{ln:sim-assign-begin}
      $t_0 \leftarrow \min\{t \mid \mathsf{Assigned}[t]=\bot\}$\;
      $v \leftarrow \textsc{PopMax}(\mathcal{H})$\;
      $\mathsf{Assigned}[t_0]\leftarrow v$\;
      $v.\mathsf{start}\leftarrow \mathsf{Offset}[t_0]$\tcp*{gas offset at which $v$ begins}\label{ln:sim-start}
      $\mathsf{Offset}[t_0]\leftarrow \mathsf{Offset}[t_0] + v.gas$\;\label{ln:sim-assign-end}
    }

    \BlankLine
    \tcp{Select earliest finishing active thread}
    $t^\star \leftarrow \arg\min_{t:\ \mathsf{Assigned}[t]\neq \bot}\ \mathsf{Offset}[t]$\;\label{ln:sim-select}
    $g_{\min} \leftarrow \mathsf{Offset}[t^\star]$\;

    \tcp{Synchronize idle threads}
    \For{$t\leftarrow 1$ \KwTo $k$}{\label{ln:sim-sync-begin}
      \If{$\mathsf{Assigned}[t]=\bot$}{
        $\mathsf{Offset}[t]\leftarrow g_{\min}$\;\label{ln:sim-sync-end}
      }
    }

    \BlankLine
    \tcp{Mark node as executed}
    $v^\star \leftarrow \mathsf{Assigned}[t^\star]$\;\label{ln:sim-exec-begin}
    $\mathsf{Assigned}[t^\star]\leftarrow \bot$\;
    $\mathsf{TotalGas}[t^\star]\leftarrow \mathsf{TotalGas}[t^\star] + v^\star.gas$\tcp*{accumulate execution cost}
    \textsc{Append}$(\mathcal{L}, v^\star)$\;\label{ln:sim-exec-end}

    \tcp{Release children}
    \ForEach{$c \in v^\star.\mathsf{Children}$}{\label{ln:sim-release-begin}
      $c.\mathsf{RemParents}\leftarrow c.\mathsf{RemParents}-1$\;
      \If{$c.\mathsf{RemParents}=0$}{
        $\textsc{Push}(\mathcal{H}, c,\; key=c.\mathsf{Path})$\;\label{ln:sim-release-end}
      }
    }
  }

  \Return{$(\mathcal{L}, \mathsf{TotalGas})$}\;
\end{algorithm}

\fi

\subsection{Identifying Hot Resources}\label{ss:hot}

Two conflict graphs should not be integrated when their transactions contend
on the same resource for much of the block: the conflicting chains cannot run
in parallel, so integration would only lengthen the critical path. We
therefore identify, in each graph, the resources that stay in active use for a
large part of the block. Execution exposes no wall-clock time at proposal
time, so ``in use'' is measured in \emph{gas}, the deterministic cost of each
transaction: the simulation of \S\ref{ss:reorder} tracks a gas offset per
thread\ifarxiv\ ($\mathsf{Offset}[t]$ in Algorithm~\ref{alg:simreorder})\fi, and records
for every transaction the offset of its thread when the transaction is
assigned\LR{ln:sim-start}.

\begin{definition}[Resource Access Span]
  For each transaction $tx$, let $g_{tx}^{\text{start}} = tx.\mathsf{start}$
  be the gas offset recorded by \ALG{alg:simreorder}{the simulation}, and
  $g_{tx}^{\text{end}} = g_{tx}^{\text{start}} + \mathrm{gas}(tx)$.
  For a resource $r$, let $\mathcal{T}_r$ be the set of transactions that
  access $r$, and define
  \begin{itemize}
    \item $g_{r}^{\text{first,start}} = \min_{tx \in \mathcal{T}_r} g_{tx}^{\text{start}}$,
          the gas offset at which the first access to $r$ begins,
    \item $g_{r}^{\text{last,end}} = \max_{tx \in \mathcal{T}_r} g_{tx}^{\text{end}}$,
          the gas offset at which the last access to $r$ ends.
  \end{itemize}
  The difference $g_{r}^{\text{last,end}} - g_{r}^{\text{first,start}}$ is
  the $r.\mathsf{span}$ of resource $r$.
\end{definition}

Let $\mathsf{MeanGas}$ be the average total gas of the $k$ simulated threads.
A resource is \emph{hot} if $r.\mathsf{span} > \mathsf{MeanGas} \cdot
\tau_{hot}$, where $\tau_{hot}$ is a configurable threshold.
Fig.~\ref{fig:algos}(d) applies the test to the example block with
$\tau_{hot}{=}0.75$: $e$, accessed at both ends of the
block, has span 4 against a threshold of 3 and is hot, while the other resources
have span 2 and are not hot.

\subsection{Identifying Balanced Conflict Graphs}\label{ss:balance}

A conflict graph whose dependency chains leave simulated threads idle is a
candidate for integration, because its idle capacity can absorb another graph's
dependency chains. An integration is worthwhile only if it fills that capacity
instead of making the critical path relatively longer. 

In Fig.~\ref{fig:algos}(c), block order
gives the two threads 6 and 2 gas in total: the dependency chain occupies one thread while
the other is idle for four units, and that idle capacity is exactly what
another graph's dependency chains could fill. The reordered schedule gives them
4 and 4: nothing is idle, so integrating more would only make the critical
path relatively longer. 

We therefore measure balance by the coefficient of variation of the
per-thread gas: with $\mathsf{TotalGas}[t]$ the gas of thread $t$, $\mu$
and $\sigma$ its mean and standard deviation over the $k$ threads, and
$\mathrm{CV}=\sigma/\mu$, $\mathsf{TotalGas}=(6,2)$ gives $\mathrm{CV}=0.5$ and
$(4,4)$ gives $\mathrm{CV}=0$.

\begin{definition}\label{def:balanced}
  Given a simulation with $k$ threads and a threshold $\tau_{\mathrm{CV}}$,
  a conflict graph is \emph{$(k,\tau_{\mathrm{CV}})$-balanced} if and only
  if $\mathrm{CV}\le\tau_{\mathrm{CV}}$.
\end{definition}

An integration of graphs $G_1,\ldots,G_m$ (\S\ref{ss:integrate}) is judged
against its parts: it must be at least as balanced as they were on average.

\begin{definition}\label{def:balanced-integration}
  An integrated conflict graph $G$ of $G_1,\ldots,G_m$ is a
  \emph{$(k,\tau_{\mathrm{CV}})$-balanced integration} if and only if
  \[
    \mathrm{CV}_G \le \mathrm{CV}_{\mathrm{base}}
      = \sum_{i=1}^{m} w_i\,\mathrm{CV}_{G_i},
    \qquad w_i = \frac{\mu_i}{\sum_{j=1}^{m}\mu_j}.
  \]
\end{definition}
\edited{This condition compares the dispersion of simulated per-thread gas loads. We use it as a proxy for the utilization benefit of integration.}

\subsection{Integrating Conflict Graphs}\label{ss:integrate}

Integration decides which of a round's graphs to integrate into one group and
how to integrate them. A greedy pass over the graphs\ALGP{alg:greedy-integration}
makes the first decision with the balance test of \S\ref{ss:balance} and the
hot resources of \S\ref{ss:hot}; a pairwise integration\ALGP{alg:merge-graphs-reorder}
 performs the second.

\heading{Choosing the groups.} \ALG{alg:greedy-integration}{The greedy} sorts the
graphs by descending CV\LR{ln:greedy-sort}, seeds a group with the
highest-CV graph\LR{ln:greedy-init}, and tries the remaining graphs
from the lowest CV upward\LR{ln:greedy-traverse}. \edited{For each candidate block $B$, the integrator compares $\mathrm{Hot}(B)$ with $\mathrm{Hot}(G)$, where $G$ is the current integrated graph, and rejects the candidate when these sets intersect. Otherwise it tentatively integrates $B$ into $G$, simulates the result, and applies the balance test\LRR{ln:greedy-test}{ln:greedy-merge}. For an accepted integration, the simulated hot-resource set becomes $\mathrm{Hot}(G)$ for the next decision.} The traversal stops as soon as the
group is itself balanced\LR{ln:greedy-break}, or when the group reaches
a cap of 10 blocks, a safeguard that rarely binds (\S\ref{sec:params}). The integrated graphs leave the list\LR{ln:greedy-remove}
 and the procedure repeats until every graph is
grouped. \edited{The test uses the integrated graph at the time of each decision. Because its hot-resource set is recomputed after integration, it does not guarantee pairwise disjointness among the original per-block hot-resource sets in the final group.}

\heading{Why this greedy.} The highest-CV graph seeds a group because it has the
most idle capacity to fill; low-CV candidates come first because they
consist of many short chains that slot into idle threads. \edited{A candidate that shares a hot resource with the current integrated graph is never integrated, since their chains on that resource would run one after the other anyway.} An integration is kept only if the group is at least as balanced as
its parts ($\mathrm{CV}_G \le \mathrm{CV}_{\mathrm{base}}$). \edited{Once the group is balanced ($\mathrm{CV}_G \le \tau_{\mathrm{CV}}$) integration stops: adding more is expected to bring little further gain while widening the optimistic window and risking more re-executions.}
\ifarxiv\BlankLine
\begin{algorithm}[t]
    \caption{\textsc{GreedyBalancedIntegration}}
    \label{alg:greedy-integration}
    \DontPrintSemicolon
    \KwIn{A set of conflict graphs $\{G_1, G_2, \ldots, G_m\}$}
    \KwOut{A list of integrated graphs $\mathcal{O}$}

    Sort the graphs in descending order of $\mathrm{CV}$ and store them in a list $L$\;\label{ln:greedy-sort}
    $\mathcal{O} \leftarrow [\,]$\;

    \While{$L \neq \emptyset$}{
        $G \leftarrow L[1]$\tcp*[r]{Initialize with highest-$\mathrm{CV}$ graph}\label{ln:greedy-init}
        $S \leftarrow \{L[1]\}$\tcp*[r]{Graphs integrated into $G$}

        \For{$i \leftarrow |L|$ \KwTo $2$}{\label{ln:greedy-traverse}
            \If{\textsc{BalancedIntegration}$(G, L[i])$}{\label{ln:greedy-test}
                $G \leftarrow \textsc{Integrate}(G, L[i])$\;\label{ln:greedy-merge}
                $S \leftarrow S \cup \{L[i]\}$\;
            }
            \If{\textsc{BalancedGraph}$(G)$}{
                \textbf{break}\;\label{ln:greedy-break}
            }
        }

        Append $G$ to $\mathcal{O}$\;
        Remove all graphs in $S$ from $L$\;\label{ln:greedy-remove}
    }
    \Return{$\mathcal{O}$}\;
\end{algorithm}
\fi

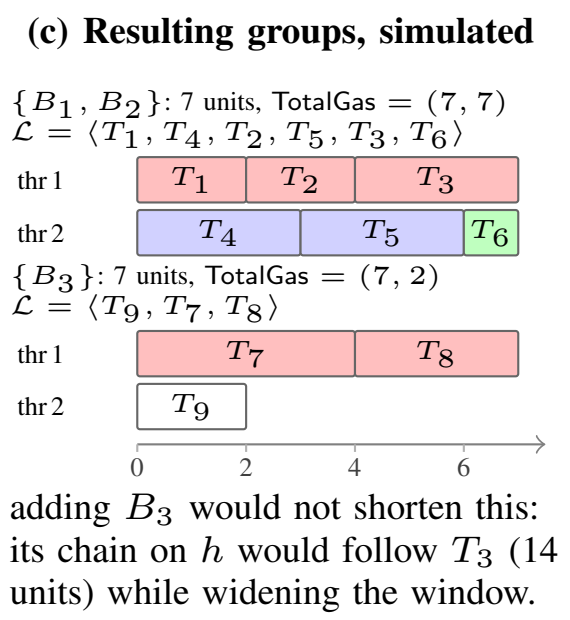
\begin{figure*}[t]
  \centering
  \begin{tikzpicture}[
      tx/.style={draw=black!70, rounded corners=1pt, inner sep=0pt,
          minimum height=16pt, minimum width=16pt, font=\scriptsize, fill=white},
      hot/.style={fill=red!25},
      hotx/.style={fill=blue!18},
      ylink/.style={fill=green!25},
      halfhy/.style={path picture={\fill[red!25] (path picture bounding box.south west) rectangle (path picture bounding box.north);
                                   \fill[green!25] (path picture bounding box.south) rectangle (path picture bounding box.north east);}},
      xdep/.style={->, >=stealth, black!70, semithick, densely dashed},
      dep/.style={->, >=stealth, black!70, semithick},
      elab/.style={font=\tiny, inner sep=1pt, fill=white},
      ptitle/.style={font=\scriptsize\bfseries, anchor=south west},
      blab/.style={font=\scriptsize, anchor=north, align=center},
      note/.style={font=\scriptsize, anchor=north west, inner sep=0pt},
      glab/.style={font=\tiny, anchor=west, inner sep=0pt},
      gbox/.style={draw=black!60, rounded corners=0.5pt, font=\tiny, inner sep=0pt},
    ]
    \begin{scope}
      \node[ptitle] at (0, 2.35) {(a) Three blocks: $\mathsf{TotalGas}$ per thread, CV, hot resources};
      \node[tx, halfhy] (T1) at (0.3, 1.75) {$T_1$};
      \node[tx, hot] (T2) at (1.25, 1.75) {$T_2$};
      \node[tx, hot] (T3) at (2.2, 1.75) {$T_3$};
      \draw[dep] (T1) -- node[elab, above] {$h$} (T2);
      \draw[dep] (T2) -- node[elab, above] {$h$} (T3);
      \node[font=\tiny, anchor=north] at (0.3, 1.45) {2};
      \node[font=\tiny, anchor=north] at (1.25, 1.45) {2};
      \node[font=\tiny, anchor=north] at (2.2, 1.45) {3};
      \node[blab] at (1.25, 0.55) {$B_1$: $\mathsf{TotalGas}=(7,0)$\\$\mathrm{CV}=1.0$, hot $\{h\}$};
      \node[tx, hotx] (T4) at (3.4, 1.75) {$T_4$};
      \node[tx, hotx] (T5) at (4.35, 1.75) {$T_5$};
      \node[tx, ylink] (T6) at (3.9, 1.0) {$T_6$};
      \draw[dep] (T4) -- node[elab, above] {$x$} (T5);
      \node[font=\tiny, anchor=north] at (3.4, 1.45) {3};
      \node[font=\tiny, anchor=north] at (4.35, 1.45) {3};
      \node[font=\tiny, anchor=west] at (4.2, 1.0) {1};
      \node[blab] at (3.9, 0.55) {$B_2$: $\mathsf{TotalGas}=(6,1)$\\$\mathrm{CV}=0.71$, hot $\{x\}$};
      \draw[xdep] (T1.south) to[out=-60, in=180] node[elab, pos=0.55, below] {$y$, added by \ifarxiv Alg.~\ref{alg:merge-graphs-reorder}\else the merge\fi} (T6.west);
      \node[tx, hot] (T7) at (5.5, 1.75) {$T_7$};
      \node[tx, hot] (T8) at (6.45, 1.75) {$T_8$};
      \node[tx]      (T9) at (7.4, 1.75) {$T_9$};
      \draw[dep] (T7) -- node[elab, above] {$h$} (T8);
      \node[font=\tiny, anchor=north] at (5.5, 1.45) {4};
      \node[font=\tiny, anchor=north] at (6.45, 1.45) {3};
      \node[font=\tiny, anchor=north] at (7.4, 1.45) {2};
      \node[blab] at (6.45, 0.55) {$B_3$: $\mathsf{TotalGas}=(7,2)$\\$\mathrm{CV}=0.56$, hot $\{h\}$};
      \draw[gray!60, densely dashed] (2.85, 2.15) -- (2.85, 0.05);
      \draw[gray!60, densely dashed] (4.95, 2.15) -- (4.95, 0.05);
      \node[note] at (0, -0.2) {gas beside each node; red: hot $h$, blue: hot $x$, green: $y$ ($T_1$ writes $h$ and $y$)};
    \end{scope}
    \begin{scope}[xshift=8.5cm]
      \node[ptitle] at (0, 2.35) {(b) \ALG{alg:greedy-integration}{the greedy}};
      \node[note, text width=5.6cm, align=left] at (0, 2.1) {%
        \setlength{\tabcolsep}{2pt}\renewcommand{\arraystretch}{1.05}\scriptsize%
        \begin{tabular}{@{}ll@{}}
          \toprule
          sort by CV & $B_1\ (1.0),\ B_2\ (0.71),\ B_3\ (0.56)$ \\
          seed $G$ & $B_1$, hot $\{h\}$ \\
          try $B_3$ & shares $h$ with $G$: skip \\
          try $B_2$ & hot $\{x\}\cap\{h\}=\emptyset$; \\
                    & $\mathrm{CV}_G{=}0\le\mathrm{CV}_{\mathrm{base}}{=}0.88$: merge \\
          re-simulate $G$ & $(7,7)$, $\mathsf{MeanGas}{=}7$, no hot left \\
                    & $\mathrm{CV}_G\le\tau_{\mathrm{CV}}$: stop \\
          output & $\{B_1,B_2\}$, then $\{B_3\}$ \\
          \bottomrule
        \end{tabular}};
    \end{scope}
    \begin{scope}[xshift=14.2cm]
      \node[ptitle] at (0, 2.35) {(c) Resulting groups, simulated};
      \pgfmathsetmacro{\u}{0.36}\pgfmathsetmacro{\xo}{0.85}\pgfmathsetmacro{\h}{0.3}
      \node[glab] at (0, 2.12) {$\{B_1, B_2\}$: 7 units, $\mathsf{TotalGas}=(7,7)$};
      \node[glab] at (0, 1.92) {$\mathcal{L}=\langle T_1,T_4,T_2,T_5,T_3,T_6\rangle$};
      \node[glab] at (0.05, 1.62) {thr\,1};\node[glab] at (0.05, 1.27) {thr\,2};
      \draw[gbox, fill=red!25] (\xo, {1.62-\h/2}) rectangle ++(2*\u, \h) node[midway] {$T_1$};
      \draw[gbox, fill=red!25] ({\xo+2*\u}, {1.62-\h/2}) rectangle ++(2*\u, \h) node[midway] {$T_2$};
      \draw[gbox, fill=red!25] ({\xo+4*\u}, {1.62-\h/2}) rectangle ++(3*\u, \h) node[midway] {$T_3$};
      \draw[gbox, fill=blue!18] (\xo, {1.27-\h/2}) rectangle ++(3*\u, \h) node[midway] {$T_4$};
      \draw[gbox, fill=blue!18] ({\xo+3*\u}, {1.27-\h/2}) rectangle ++(3*\u, \h) node[midway] {$T_5$};
      \draw[gbox, fill=green!25] ({\xo+6*\u}, {1.27-\h/2}) rectangle ++(\u, \h) node[midway] {$T_6$};
      \node[glab] at (0, 0.97) {$\{B_3\}$: 7 units, $\mathsf{TotalGas}=(7,2)$};
      \node[glab] at (0, 0.77) {$\mathcal{L}=\langle T_9,T_7,T_8\rangle$};
      \node[glab] at (0.05, 0.47) {thr\,1};\node[glab] at (0.05, 0.12) {thr\,2};
      \draw[gbox, fill=red!25] (\xo, {0.47-\h/2}) rectangle ++(4*\u, \h) node[midway] {$T_7$};
      \draw[gbox, fill=red!25] ({\xo+4*\u}, {0.47-\h/2}) rectangle ++(3*\u, \h) node[midway] {$T_8$};
      \draw[gbox, fill=white] (\xo, {0.12-\h/2}) rectangle ++(2*\u, \h) node[midway] {$T_9$};
      \draw[->, gray] (\xo, -0.13) -- ({\xo+7.5*\u}, -0.13);
      \foreach \i in {0,2,4,6} {\draw[gray!60] ({\xo+\i*\u}, -0.13) -- ++(0,-0.05); \node[font=\tiny, gray!50!black] at ({\xo+\i*\u}, -0.28) {\i};}
      \node[note, text width=3.8cm, align=left] at (0, -0.45) {adding $B_3$ would not shorten this: its chain on $h$ would follow $T_3$ (14 units) while widening the window.};
    \end{scope}
  \end{tikzpicture}
  \caption{Integration of a three-block round ($k{=}2$,
    $\tau_{hot}{=}1.5$, $\tau_{\mathrm{CV}}{=}0.5$). (a)~Per-block
    simulation: $B_1$ and $B_3$ each has a hot resource $h$, $B_2$ has a hot resource 
    $x$. (b)~The greedy integration starts with $B_1$, skips $B_3$
    because it shares the hot resource $h$, and integrates $B_2$, whose hot
    resource differs. \ALG{alg:merge-graphs-reorder}{the integration} adds the
    dashed edge $T_1 \to T_6$. (c)~The two groups take 14 units in sequence
    against 20 for the three blocks alone; adding $B_3$ to the group would
    serialize both chains on $h$ and gain nothing.}
  \label{fig:integration}
\end{figure*}

Fig.~\ref{fig:integration} runs the greedy on a three-block round.

\heading{Integrating two graphs.} Integrating $G_2$ into $G_1$ treats $G_2$'s
block as if it followed $G_1$'s: every cross-block conflict becomes an edge
from the $G_1$ transaction to the $G_2$ transaction, and pairs that do not
conflict get no edge and may run in parallel. The per-resource maps make this
a local operation: for a resource $r$, only $G_1$'s last writer and the
readers since that write can conflict with $G_2$, and only $G_2$'s heads
(its accessors before its first write of $r$) and that first writer can
conflict with $G_1$. In Fig.~\ref{fig:integration}, integrating $B_2$ after $B_1$ adds the
single edge $T_1 \to T_6$: $T_1$ is $G_1$'s last writer of $y$ and $T_6$
is $G_2$'s only accessor of it.

\ALG{alg:merge-graphs-reorder}{The integration} connects exactly those pairs,
resource by resource\LR{ln:int-foreach}: $G_1$'s tail to every
head of $r$ in $G_2$, a RAW edge to each reader before $G_2$'s first write
and a WAW edge to that writer\LRR{ln:int-tail-begin}{ln:int-tail-end}, and $G_1$'s tail
readers to $G_2$'s first writer (WAR\LRc{ln:int-war}). The maps are
then integrated so that $G$ can be integrated further: the heads of $r$ stay
$G_1$'s where $G_1$ wrote $r$ and gain $G_2$'s otherwise\LR{ln:int-notail-begin}; the tail and tail readers become $G_2$'s
where $G_2$ wrote $r$\LR{ln:int-tail-update}, and $G_2$'s readers
join $G_1$'s tail readers otherwise\LR{ln:int-notail-end}.

Finally the integrated graph is simulated once more, exactly as a single
block's graph is (\S\ref{ss:reorder}\LRc{ln:int-reorder}). The
simulation recomputes the transaction order, CV and hot resources in the integrated graph.
\ifarxiv\begin{algorithm}[t]
\caption{\textsc{IntegrateGraphsAndReorder}: Integrate $G_1$ and $G_2$ and Reorder Transactions}
\label{alg:merge-graphs-reorder}
\DontPrintSemicolon
\KwIn{$G_1=(V_1,E_1)$, $G_2=(V_2,E_2)$, each with its maps $\mathsf{Head}$, $\mathsf{Tail}$, $\mathsf{TailReaders}$ (a resource a graph never touched reads as $\emptyset$, $\bot$, $\emptyset$); number of threads $k$}
\KwOut{Reordered list of transactions $\mathcal{L}$}

$V \leftarrow V_1 \cup V_2$;\ $E \leftarrow E_1 \cup E_2$\;
$(\mathsf{Head}, \mathsf{Tail}, \mathsf{TailReaders}) \leftarrow (G_1.\mathsf{Head}, G_1.\mathsf{Tail}, G_1.\mathsf{TailReaders})$\;

\BlankLine
\tcp{Connect every cross-graph conflict on each resource $G_2$ accesses}
\ForEach{$r$ accessed in $G_2$}{\label{ln:int-foreach}
  \uIf{$\mathsf{Tail}[r]\neq\bot$}{
    \tcp{$G_1$ wrote $r$: RAW edges to $G_2$'s readers before its first write, WAW edge to that writer}
    \ForEach{$v \in G_2.\mathsf{Head}[r]$}{\label{ln:int-tail-begin}
      $\textsc{AddEdge}(E,\ \mathsf{Tail}[r],\ v)$\;\label{ln:int-tail-end}
    }
  }
  \Else{
    $\mathsf{Head}[r]\leftarrow \mathsf{Head}[r]\ \cup\ G_2.\mathsf{Head}[r]$\;\label{ln:int-notail-begin}
  }
  \uIf{$G_2.\mathsf{Tail}[r]\neq\bot$}{
    \tcp{$G_2$ wrote $r$: WAR edges from $G_1$'s readers since its last write to $G_2$'s first writer}
    $w \leftarrow$ the node of $G_2.\mathsf{Head}[r]$ that writes $r$\;
    \lForEach{$u \in \mathsf{TailReaders}[r]$}{$\textsc{AddEdge}(E,\ u,\ w)$}\label{ln:int-war}
    $\mathsf{Tail}[r]\leftarrow G_2.\mathsf{Tail}[r]$\\ $\mathsf{TailReaders}[r]\leftarrow G_2.\mathsf{TailReaders}[r]$\;\label{ln:int-tail-update}
  }
  \Else{
    $\mathsf{TailReaders}[r]\leftarrow \mathsf{TailReaders}[r]\ \cup\ G_2.\mathsf{TailReaders}[r]$\;\label{ln:int-notail-end}
  }
}

\BlankLine
\tcp{After integration, reorder transactions using the transaction reordering algorithm}
$G \leftarrow \big((V,E),\mathsf{Head},\mathsf{Tail},\mathsf{TailReaders}\big)$\;
$(\mathcal{L}, \mathsf{TotalGas}) \leftarrow \textsc{TxnReorder}(G,\ k)$\;\label{ln:int-reorder}

\Return{$(\mathcal{L}, \mathsf{TotalGas})$}\;
\end{algorithm}
\fi

\subsection{Graph-Aware OCC Engine}\label{ss:occ}

To enable parallel execution, we deploy a graph-aware OCC engine\ALGP{alg:occ-dg}, a variant of Block-STM that follows the conflict graph during execution. Transaction indices are positions in the order fixed by the simulation (\S\ref{ss:reorder}, \S\ref{ss:integrate}), so multi-version reads and validation follow the fixed order. Transactions are executed in parallel, and the graph decides when a transaction becomes eligible.

Correctness rests on \emph{sequential validation}, not on the graph. Because the predicted read and write sets may be inaccurate, each executed transaction is validated in increasing index order: validation re-reads every resource the transaction observed from the multi-version state, and if a lower-indexed transaction has since written it, the read is invalid and the transaction is re-executed as a new incarnation\LR{ln:occ-abort}. Every re-execution, and every incarnation that writes a resource its predecessor did not, puts all higher-indexed transactions back into the validation queue\ifarxiv\ (Lines~\ref{ln:occ-abort} and~\ref{ln:occ-newres})\fi, so no transaction commits having read a version that a lower-indexed transaction later changed. Validation is prioritized over execution.

We maintain an execution task pool $\mathsf{ETask}$, which contains transactions whose RAW parents have all executed and that are ready for execution\LR{ln:occ-etask-init}. After a transaction is executed, a corresponding validation task is created and added to the validation task pool $\mathsf{VTask}$\LR{ln:occ-exec}. At the same time, any child transactions whose dependencies have now been satisfied are inserted into $\mathsf{ETask}$\LR{ln:occ-children}. When selecting tasks from $\mathsf{ETask}$ and $\mathsf{VTask}$, we prioritize tasks with lower transaction indices\LR{ln:occ-pick}.

The key difference from Block-STM is how $\mathsf{ETask}$ is maintained: Block-STM starts with every transaction in it, whereas our engine admits only transactions whose RAW parents in $G$ have all executed\LR{ln:occ-etask-init}. Only RAW parents are waited for: sequential validation makes the other two kinds of edge unnecessary at execution time. A write by a higher-indexed transaction is never visible to a lower-indexed reader, whose reads resolve to versions below its own index\LR{ln:occ-read}. If two writers write different values, the one with higher index wins. And when a re-execution happens, the transaction and all transactions with higher indices should be validated again. A missing RAW parent, by contrast, means a read that validation will fail, so admitting a transaction before its RAW parents have executed should not be allowed.

\section{Evaluation}
\label{sec:evaluation}

\ifarxiv\begin{algorithm}[t]
\caption{\textsc{ParExecOCCDG}: Parallel Execution with OCC + Conflict Graph}
\label{alg:occ-dg}
\DontPrintSemicolon
\KwIn{Transaction list $\mathcal{L}=\langle tx_1,\dots,tx_n\rangle$; conflict graph $G$; number of threads $k$}
\KwOut{Final execution results for all transactions}
$\mathsf{ETask}\leftarrow \{tx_i \mid tx_i \text{ has no unexecuted RAW parents in } G\}$\label{ln:occ-etask-init}\tcp*{ready-to-execute}
$\mathsf{VTask}\leftarrow \emptyset$\tcp*{executed but not validated}
\lFor{$i\leftarrow 1$ \KwTo $n$}{$\mathsf{Inc}[i]\leftarrow 0$}
\tcp{Multi-version state used by \textsc{Execute}}
$\textsc{Read}(tx_i, r)$ returns the version of $r$ written by the highest-indexed $tx_j$ with $j<i$, or the pre-round value;\ $\textsc{Write}(tx_i, r, v)$ creates version $(i,\mathsf{Inc}[i])$\;\label{ln:occ-read}
\While{$\mathsf{ETask}\neq \emptyset \ \vee\ \mathsf{VTask}\neq \emptyset$}{
  $(\mathsf{kind}, i)\leftarrow \textsc{PickTask}(\mathsf{VTask},\mathsf{ETask})$\label{ln:occ-pick}\tcp*{validation before execution; lowest index first}
  \uIf{$\mathsf{kind}=\textsc{Validate}$}{
    \If{$\neg\textsc{Validate}(tx_i,\ \mathsf{Inc}[i])$}{
      $\mathsf{Inc}[i]\leftarrow \mathsf{Inc}[i]+1$;\ $\mathsf{ETask}\leftarrow \mathsf{ETask}\cup\{i\}$\tcp*{abort and re-execute}\label{ln:occ-abort}
      \For{$j\leftarrow i+1$ \KwTo $n$}{\lIf{$j \notin \mathsf{ETask}$}{$\mathsf{VTask}\leftarrow \mathsf{VTask}\cup\{j\}$}}
    }
  }
  \Else{
    $(R,W)\leftarrow \textsc{Execute}(tx_i,\ \mathsf{Inc}[i])$;\ $\mathsf{VTask}\leftarrow \mathsf{VTask}\cup\{i\}$\;\label{ln:occ-exec}
    \ForEach{$j \in tx_i.\mathsf{Children}$}{\label{ln:occ-children}
      \lIf{$\textsc{RawParentsExecuted}(j)$}{$\mathsf{ETask}\leftarrow \mathsf{ETask}\cup\{j\}$}
    }
    \If{$\textsc{WritesNewResource}(i,\ W)$}{\label{ln:occ-newres}
      \For{$j\leftarrow i+1$ \KwTo $n$}{
        \lIf{$j \notin \mathsf{ETask}$}{$\mathsf{VTask}\leftarrow \mathsf{VTask}\cup\{j\}$}
      }
    }
  }
}
\end{algorithm}
\fi

Our evaluation answers four questions. 
\begin{enumerate}
\item How much faster does \sysname execute a round than single-block parallel execution engines, and than the
inter-block baselines that execute the whole round as one optimistic window (\S\ref{sec:speedup})?
\item Under which conflict pattern and parameters does integration pay off (\S\ref{sec:when})?
\item Why does it work, in terms of aborts (\S\ref{sec:why})?
\item \edited{And what does each stage cost, including the stages that single-block engines do not have, and how much of that cost can a validator's pipeline hide (\S\ref{sec:e2e})?}
\end{enumerate}

\subsection{Methodology}
\label{sec:methodology}

\heading{Implementation.}
\label{sec:implementation}
\sysname is about 6{,}500 lines of Rust on top of pevm~\cite{pevm}, an
open-source Block-STM-style parallel EVM executor. On top of it, we implement conflict-graph
construction, the reordering simulation, the greedy integrator and the
graph-aware OCC engine. 

\heading{Testbed.}
One Amazon EC2 \texttt{c4.8xlarge} instance: two Intel Xeon E5-2666~v3
sockets (18 cores, 36 hardware threads, 2.9\,GHz), 58\,GiB of DRAM,
Ubuntu 24.04.

\heading{Baselines.}
\label{sec:baselines}
All engines share one code base, one \texttt{revm} and one in-memory
state backend, and every hint-based engine consumes the access sets of the
same sequential pre-execution, so no baseline receives better hints than
\sysname.
\begin{itemize}\setlength{\itemsep}{0pt}
  \item \emph{Sequential}: single-threaded \texttt{revm}.
  \item \emph{Block-STM}: each block executed with pevm's optimistic
        multi-version engine.
  \item \emph{Graph-aware OCC}: each block executed with the engine of
        \S\ref{ss:occ} on the block's own conflict graph.
  \item \emph{Concatenated Block-STM} and \emph{concatenated graph OCC}:
        all blocks of a round integrated, in order, into one block and run with
        unmodified Block-STM, or with the graph-aware engine on a conflict
        graph built over the whole round. These are the trivial
        inter-block baselines: the largest possible optimistic window, no
        reordering and no hot-resource isolation.
  \item \emph{Vegeta}~\cite{vegeta}, following its paper's
        Algorithms~1--3: transactions are re-sequenced so that the longest
        per-key dependency chains run first, then partitioned along the
        speculative dependency DAG into batches whose transactions run in
        parallel against the state at the start of the batch. A
        transaction whose actual access set differs from the speculated
        one is executed sequentially at the end. It has no multi-version
        store and is intra-block only.
  \item \emph{\sysname}: the per-block graphs integrated into groups
        (\S\ref{ss:integrate}), each group executed with the graph-aware OCC
        engine.
\end{itemize}

\begin{figure}[t]
  \tiny
\begin{verbatim}
mapping(bytes32 => uint256) public state;

function execute(bytes32[] calldata reads,
                 bytes32[] calldata writes,
                 uint256 target) external {
  uint256 count = 0;
  while (count < target) {
    for (uint i = 0; i < reads.length
                     && count < target; i++) {
      state[reads[i]];       count++;  // load
    }
    for (uint i = 0; i < writes.length
                     && count < target; i++) {
      state[writes[i]] += 1; count++;  // increment
    }
  }
}
\end{verbatim}
  \caption{The simulator contract behind the synthetic and artificial
    workloads. A transaction is one call to \texttt{execute}: the two key
    lists fix which slots it touches and \texttt{target} fixes how much
    work it does.}
  \label{fig:simulator}
\end{figure}

\begin{table}[t]
  \centering\footnotesize\setlength{\tabcolsep}{3pt}
  \caption{Default configuration.}
  \label{tab:defaults}
  \begin{tabular}{@{}ll@{}}
    \toprule
    Parameter & Default \\
    \midrule
    Workers $k$ (all engines) & 8 \\
    Hot-resource threshold $\tau_{hot}$ & 1.5 \\
    Balance threshold $\tau_{\mathrm{CV}}$ & 0.5 \\
    Max blocks per group & 10 \\
    Number of rounds & 100 blocks; 150 rounds \\
    \bottomrule
  \end{tabular}
\end{table}

\begin{table*}[t]
  \centering\footnotesize
  \caption{Comparison between engines. Throughput,
    speedup, per-block latency, and re-executions}
  \label{tab:absolute}
  \begin{tabular}{l rrrrr rrrrr}
    \toprule
    & \multicolumn{5}{c}{Real Ethereum} & \multicolumn{5}{c}{Synthetic} \\
    \cmidrule(lr){2-6}\cmidrule(lr){7-11}
    Engine & tx/s & speedup & p50 ms & p99 ms & re-exec/tx
           & tx/s & speedup & p50 ms & p99 ms & re-exec/tx \\
    \midrule
    Sequential              & 17{,}322 & 1.00$\times$ & 8.91 & 15.00 & --
                            & 12{,}097 & 1.00$\times$ & 12.50 & 22.77 & -- \\
    Block-STM               & 32{,}823 & 1.89$\times$ & 4.58 & 9.76 & 0.55
                            & 24{,}381 & 2.02$\times$ & 6.06 & 17.04 & 0.73 \\
    Concatenated Block-STM  & 41{,}523 & 2.40$\times$ & 3.67 & 8.45 & 0.88
                            & 34{,}186 & 2.83$\times$ & 4.41 & 13.17 & 1.16 \\
    Concatenated graph OCC  & 37{,}501 & 2.17$\times$ & 4.05 & 8.55 & 0.55
                            & 36{,}181 & 2.99$\times$ & 4.18 & 12.97 & 0 \\
    Graph-aware OCC         & 28{,}805 & 1.66$\times$ & 5.35 & 10.45 & 0.27
                            & 25{,}156 & 2.08$\times$ & 5.86 & 16.80 & 0 \\
    Vegeta                  & 22{,}966 & 1.33$\times$ & 6.81 & 12.26 & --
                            & 16{,}558 & 1.37$\times$ & 9.13 & 20.04 & -- \\
    \sysname                & \textbf{42{,}757} & \textbf{2.47$\times$} & \textbf{3.73} & \textbf{5.50} & 0.31
                            & 27{,}627 & 2.28$\times$ & 5.36 & 15.19 & 0 \\
    \bottomrule
  \end{tabular}
\end{table*}

\heading{Workloads.}
\begin{itemize}\setlength{\itemsep}{0pt}
  \item \emph{Real Ethereum}: 15{,}000 mainnet blocks in three contiguous
        5{,}000-block segments, 16{,}774{,}645--16{,}779{,}644 (March
        2023), 18{,}581{,}726--18{,}586{,}725 (November
        2023) and 19{,}557{,}289--19{,}562{,}288 (April 2024), about 2.4\,M transactions in all. For every block an
        archive node supplies the header, the full transaction objects
        and the prestate value of every account and storage slot the
        block touches. 
        
        EVM execution is state-dependent: a transaction scheduled ahead of
        one it followed on mainnet observes a different state, may branch
        to other keys, and causes aborts that are artifacts of replay
        rather than of scheduling. To avoid fee failures, account balances are raised so that no
        reordering fails a fee check. To avoid nonce failures, the engines that reorder
        renumber each sender's nonces in the executed order. For the same reason this is the
        workload in which hints are imperfect: pre-execution records each
        transaction's access sets in mainnet order, and a transaction
        executed in another order can touch keys its hint did not
        predict. The next two workloads are not subject to these effects.
  \item \emph{Synthetic}: a synthetic version of the real blocks that keeps their
        conflict topology and execution-time distribution but is
        insensitive to order, in the manner of~\cite{chiron}. We performed a one-time
        sequential replay of every block against its prestate, recording
        each transaction's read set and write set (one key for each storage
        slot or account balance),
        its wall-clock execution time, and its sender. 
        
        One simulator contract (Fig.~\ref{fig:simulator}) holds a single
        \texttt{mapping(bytes32 => uint256)} and exposes one function,
        \texttt{execute(reads, writes, target)}, that cycles over the two
        key lists, one load per read key and one increment per write key,
        until exactly \texttt{target} storage accesses have been
        performed. In order to make the execution time match those of the real Ethereum transactions, each synthetic transaction calls this function with
        the recorded key lists and
        $\texttt{target}=\lfloor t_{\mathrm{exec}}/T_{\mathrm{SLOAD}}\rfloor$,
        where $T_{\mathrm{SLOAD}}\approx0.48\,\mu$s is the measured cost of
        one storage access on our testbed.
        
        Each distinct sender
        is mapped to one fresh pre-funded account that is reused across
        all blocks, which can preserve nonce chains across blocks. The
        keys a transaction touches are fixed by its calldata, so hints
        are exact. 

  \item \emph{Artificial}: the same simulator contract with generated
        access sets whose conflict is controlled by two parameters,
        $\rho_{\mathrm{intra}}$ for conflict inside a block and
        $\rho_{\mathrm{inter}}$ for conflict across blocks, so that
        \S\ref{sec:when} can vary conflict patterns while everything else
        is fixed. Every transaction writes three cold keys drawn uniformly
        from a pool of 10{,}000, and every block is assigned one
        \emph{hot key} from a set $\{h_1,\dots,h_H\}$. Because every
        write is an increment, the transactions that write the same hot
        key form a serial chain, and the two parameters set the shape of
        that chain:
        \begin{itemize}[nosep,leftmargin=1.5em]
          \item $\rho_{\mathrm{intra}}$ controls intra-block conflict: the
                first $\rho_{\mathrm{intra}}$ percent of a block's
                transactions also write the block's hot key, and the
                block is then shuffled so that they are spread across it.
                It sets how many transactions of one block chain on the
                same key.
          \item $\rho_{\mathrm{inter}}$ controls inter-block conflict:
                consecutive blocks are taken in groups of
                $H{-}1{+}\rho_{\mathrm{inter}}$; the first
                $\rho_{\mathrm{inter}}$ blocks of a group all use $h_1$
                and the remaining $H{-}1$ blocks use $h_2,\dots,h_H$, one
                each. It sets how many consecutive blocks chain on the
                same key.
        \end{itemize}
        The cold keys add only a random, unstructured background of
        conflicts; per-transaction work is fixed at 100 storage accesses
        and every transaction has a fresh sender, so the conflict
        structure is the only variable.
\end{itemize}

\heading{Defaults and metrics.}
\label{sec:params}
Table~\ref{tab:defaults} lists the configuration every experiment uses
unless the text states otherwise. The worker count is the one at which
every engine peaks on this machine (\S\ref{sec:speedup}); the cap on group size rarely binds, since groups average 4.2 blocks on real Ethereum and 1.3 on
the synthetic workload. Speedup is over single-threaded sequential
\texttt{revm} on the same blocks, re-measured alongside each parallel run,
and covers the execution phase only; \S\ref{sec:e2e} charges the
preparatory stages. On real Ethereum the 15 rounds in which the Block-STM
engine abandoned an optimistic window and re-executed it sequentially (an
implementation fallback) are excluded from every statistic, leaving 135
rounds.

\subsection{Execution Speedup}
\label{sec:speedup}
\label{sec:absolute}

This subsection answers the first question, how much faster \sysname
executes a round than the per-block engines and the concatenated round;
Table~\ref{tab:absolute} measures every engine of \S\ref{sec:baselines}
in one run over the same rounds.

\heading{Comparing with per-block engines.}
Three of the engines in Table~\ref{tab:absolute} execute one block at a
time. On real Ethereum, Block-STM reaches 1.89$\times$ speedup. Graph-aware OCC
reaches only 1.66$\times$, although it has a conflict graph that
Block-STM lacks. The graph is a prediction made by pre-executing the
block, and a transaction that actually runs can touch keys it did not
touch then, or skip keys it did. A missing edge costs
a re-execution, and a spurious edge makes a transaction wait for a parent
it does not need. Vegeta is the slowest at 1.33$\times$ because it runs one batch in parallel
and waits for all of it to finish before starting the next. Thus, in Vegeta, a block
with long dependency chains needs many batches, and at the end of each batch most workers sit idle. 
\sysname uses the same imperfect graphs as graph-aware OCC, but reaches
2.47$\times$, 31\% more than Block-STM and faster than it on 127 of the 135 rounds, and its slow blocks are faster
too: the p99 per-block latency is 5.50\,ms against 9.76\,ms for
Block-STM. The difference is that \sysname schedules multiple blocks
together, so workers always have more independent transactions to pick
from than one block contains. 

On the synthetic workload the hints are exact. Block-STM reaches 2.02$\times$, graph-aware OCC
2.08$\times$, Vegeta 1.37$\times$ and \sysname 2.28$\times$. With exact hints graph-aware OCC never re-executes a transaction. Block-STM, however, still
aborts and retries on its own (0.73 re-executions per transaction).
\sysname again adds the
inter-block parallelism that no per-block engine provides.

\begin{table}[t]
  \centering\scriptsize\setlength{\tabcolsep}{4pt}
  \caption{Influence of the number of blocks per round.}
  \label{tab:roundsize}
  \begin{tabular}{l rrr}
    \toprule
    Blocks per round $B$ & 20 & 50 & 100 \\
    \midrule
    Block-STM              & 1.83 & 1.86 & 1.89 \\
    Concatenated Block-STM & \textbf{2.31} & \textbf{2.32} & 2.40 \\
    \sysname               & 2.19 & \textbf{2.32} & \textbf{2.47} \\
    \bottomrule
  \end{tabular}
\end{table}

\heading{Comparing with the concatenated round.}
The strongest baseline is also the simplest: treat the 100 blocks of a
round as one big block and run Block-STM on it. This spreads every fixed
per-block cost over the round and gives the optimistic engine the largest
possible pool of transactions. It is \sysname with no limit on group
size and no hot-resource test, and \sysname does not beat it everywhere.

The concatenated round wins in two situations. First, when the hints
are exact. On the synthetic workload no graph-aware OCC engine ever
re-executes a transaction, and graph-aware OCC run over the concatenated round
reaches 2.99$\times$ speedup against 2.28$\times$ speedup for \sysname. On the
artificial workload it is the best engine at every conflict level
(\S\ref{sec:when}). Second, when the number of blocks per round is small. Over the real Ethereum workload it leads
at 20 blocks per round, 2.31$\times$ against 2.19$\times$, ties at 50
and loses at 100 (Table~\ref{tab:roundsize}). A 20-block window
already can fully utilize the eight workers, so round
size changes little for the concatenated window. For \sysname it changes
how many balanced integrations the integrator can find, since it only
integrates blocks that pass the hot-resource and balance tests, 
where we observed that the blocks per group grow
from 2.9 blocks on average at 20 blocks per round to 4.2 at 100. 

\sysname wins where one large optimistic window becomes fragile, which happens 
on the real Ethereum workload for two reasons. First, the hints are imperfect,
so conflicts between transactions are discovered only at validation time. In
the concatenated round, a transaction that reads a key written by an
earlier block of the round can run before that write is done. Then it fails 
validation and re-executes, and each re-execution can invalidate the
transactions that read its results. Based on our measurements, this happens 0.88 times per
transaction, 61\% more than for Block-STM running block by block, which
does not validate across blocks. \sysname integrates only blocks that pass the
hot-resource and balance tests into one group and executes groups one
after another, so a conflict between two groups does not need validation.
Inside a group, the integrated graph holds a transaction back
until its RAW parents have been executed. It turns out to re-execute 0.31 times per
transaction and runs 3\% faster (2.47$\times$ against 2.40$\times$) than the
concatenated round with block-STM.
Giving the concatenated round a conflict graph does not fix this. It
cuts the re-executions by a third but has a lower speedup of 2.17$\times$,
because over a 15{,}000-transaction group the predicted edges make too
many transactions wait. Second, \sysname performs more stably across
rounds. Its p99 per-block latency is 5.50\,ms against 8.45\,ms for the
concatenated round (p90 4.35\,ms against 5.40\,ms). The slow rounds
are the ones with many conflicts between blocks. In one large window
every such conflict adds re-executions that can spread through the
whole window, so a highly conflicted round takes much longer than a typical one.
\edited{\sysname's groups are small, and each block is integrated only if it shares no hot resource with the integrated graph, which limits the impact of a bad block to its own group.}

\begin{table}[t]
  \centering\scriptsize\setlength{\tabcolsep}{3pt}
  \caption{Per-round speedup with real-Ethereum workloads.}
  \label{tab:worst}
  \begin{tabular}{l rrr}
    \toprule
    & \multicolumn{3}{c}{speedup over sequential, per round} \\
    \cmidrule(lr){2-4}
    Engine & slowest round & slowest tenth & slowest round/median \\
    \midrule
    Block-STM              & 1.36 & 1.63 & 0.69 \\
    Concatenated Block-STM & 1.51 & 1.93 & 0.61 \\
    Concatenated graph OCC & 1.43 & 1.92 & 0.67 \\
    Graph-aware OCC        & 1.27 & 1.47 & 0.74 \\
    Vegeta                 & 0.84 & 1.24 & 0.63 \\
    \sysname               & \textbf{1.86} & \textbf{2.09} & \textbf{0.81} \\
    \bottomrule
  \end{tabular}
\end{table}

\begin{table}[t]
  \centering\scriptsize\setlength{\tabcolsep}{3pt}
  \caption{The five rounds on which the concatenated round is slowest: speedup
    over sequential and re-executions per transaction.}
  \label{tab:worstfive}
  \begin{tabular}{l rr rr rr}
    \toprule
    & \multicolumn{2}{c}{Block-STM} & \multicolumn{2}{c}{Concatenated} & \multicolumn{2}{c}{\sysname} \\
    \cmidrule(lr){2-3}\cmidrule(lr){4-5}\cmidrule(lr){6-7}
    Round starting block & speedup & re-exec & speedup & re-exec & speedup & re-exec \\
    \midrule
    18{,}582{,}826 & 1.36 & 0.54 & 1.51 & 0.90 & \textbf{3.11} & \textbf{0.33} \\
    18{,}583{,}826 & 1.37 & 0.72 & 1.59 & 1.22 & \textbf{2.70} & \textbf{0.36} \\
    18{,}586{,}026 & 1.54 & 0.55 & 1.69 & 1.15 & \textbf{2.90} & \textbf{0.39} \\
    18{,}586{,}226 & 1.61 & 0.64 & 1.73 & 1.06 & \textbf{2.46} & \textbf{0.41} \\
    18{,}582{,}026 & 1.51 & 0.67 & 1.78 & 1.18 & \textbf{2.60} & \textbf{0.44} \\
    \bottomrule
  \end{tabular}
\end{table}

\heading{What each design buys.}
When the hints are exact, the concatenated round has the higher average
throughput. What \sysname buys instead is a per-round execution time that stays
stable no matter how conflicted the round is, and that is what
\emph{congestion control} in our title means. An optimistic window with
no size limit is fast on average but slower on a round with many
conflicts between blocks. 

Table~\ref{tab:worst} looks at the 135
real-Ethereum rounds one by one. The concatenated round's slowest round runs at 1.51$\times$ speedup
against 1.86$\times$ for \sysname, and the slowest tenth of its rounds
run at 1.93$\times$ or less, against 2.09$\times$. Its slow rounds are
also further below its typical ones: its slowest round reaches 61\% of
its median round's speedup, against 81\% for \sysname.

Table~\ref{tab:worstfive} lists the five rounds on which the
concatenated round is slowest: it runs at 1.51--1.78$\times$ with
0.9--1.2 re-executions per transaction, Block-STM is no better, and
\sysname runs at 2.5--3.1$\times$ with 0.3--0.4 re-executions. The reason is that \sysname fixes its groups
before execution, \edited{and admits a block to a group only if it shares no hot resource with the group as integrated so far, so} a round with many
conflicts between blocks gets smaller groups instead of more
re-executions. A validator must execute every round before the next, 
so the slow rounds determine its pace. \sysname gives up some average
throughput for a stable per-round time. 

\subsection{When Does Integration Pay Off?}
\label{sec:when}
\label{sec:artificial}

This subsection answers the second question: under which conflict
pattern and parameter settings integration pays off. The artificial workload of
\S\ref{sec:methodology} allows us to customize the conflict pattern. We sweep
$\rho_{\mathrm{inter}} \in \{2,4,6,8\}$ and
$\rho_{\mathrm{intra}} \in \{10,20,30,40\}\%$ over 100 rounds of 50
blocks of 100 transactions, with a hot set of $H{=}5$ keys and the
default \sysname parameters. Block-STM and graph-aware OCC execute each block on
its own, so their speedup depends on $\rho_{\mathrm{intra}}$ alone.

\pgfplotsset{artaxis/.style={
      width=0.9\columnwidth, height=2.7cm,
      xtick={10,20,30,40},
      xlabel={\small $\rho_{\mathrm{intra}}$ (\% txs touching hot key, intra-block)},
      ylabel={\small Speedup},
      xlabel style={font=\bfseries}, ylabel style={font=\bfseries},
      ymin=1.8, ymax=6.5,
      tick label style={font=\small\bfseries},
      axis line style={gray!60},
      ymajorgrids=true, grid style={dotted, gray!50},
      legend style={font=\tiny, draw=none,
          at={(0.5,1.03)}, anchor=south, column sep=5pt},
      legend columns=2,
      legend cell align=left,
    }}

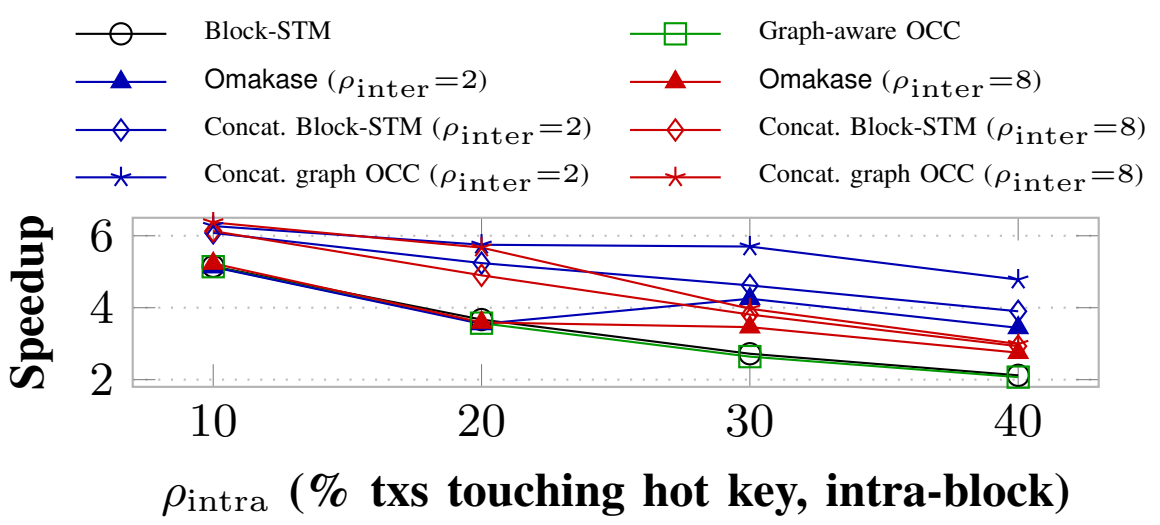
\begin{figure}[t]
  \centering
  \begin{tikzpicture}
    \begin{axis}[artaxis]
      \addplot[mark=o, color=black, thin]
      coordinates {(10,5.15) (20,3.67) (30,2.72) (40,2.12)};
      \addplot[mark=square, color=green!60!black, thin]
      coordinates {(10,5.14) (20,3.57) (30,2.64) (40,2.07)};
      \addplot[mark=triangle*, color=blue!70!black, thin]
      coordinates {(10,5.13) (20,3.55) (30,4.25) (40,3.44)};
      \addplot[mark=triangle*, color=red!80!black, thin]
      coordinates {(10,5.23) (20,3.59) (30,3.46) (40,2.75)};
      \addplot[mark=diamond, color=blue!70!black, thin]
      coordinates {(10,6.08) (20,5.24) (30,4.62) (40,3.90)};
      \addplot[mark=diamond, color=red!80!black, thin]
      coordinates {(10,6.13) (20,4.90) (30,3.81) (40,2.93)};
      \addplot[mark=star, color=blue!70!black, thin]
      coordinates {(10,6.27) (20,5.75) (30,5.70) (40,4.78)};
      \addplot[mark=star, color=red!80!black, thin]
      coordinates {(10,6.37) (20,5.67) (30,3.97) (40,2.99)};
      \legend{Block-STM, Graph-aware OCC,
        \sysname\ ($\rho_{\mathrm{inter}}{=}2$), \sysname\ ($\rho_{\mathrm{inter}}{=}8$),
        Concat.\ Block-STM ($\rho_{\mathrm{inter}}{=}2$), Concat.\ Block-STM ($\rho_{\mathrm{inter}}{=}8$),
        Concat.\ graph OCC ($\rho_{\mathrm{inter}}{=}2$), Concat.\ graph OCC ($\rho_{\mathrm{inter}}{=}8$)}
    \end{axis}
  \end{tikzpicture}
  \caption{Artificial workload, hot set $H{=}5$. Block-STM and graph-aware
    OCC depend only on $\rho_{\mathrm{intra}}$; the three inter-block
    engines are shown for $\rho_{\mathrm{inter}}{=}2$ (blue) and $\rho_{\mathrm{inter}}{=}8$ (red).}
  \label{fig:artificial}
\end{figure}

\pgfplotsset{tauaxis/.style={
      width=0.31\textwidth, height=3.2cm,
      xlabel style={font=\small\bfseries}, ylabel style={font=\small\bfseries},
      tick label style={font=\scriptsize},
      axis line style={gray!60},
      ymajorgrids=true, grid style={dotted, gray!50},
    }}

\begin{figure*}[t]
  \centering
  \begin{tikzpicture}
    \begin{axis}[tauaxis, xmode=log, xlabel={$\tau_{hot}$}, ylabel={Speedup},
        xmin=0.42, xmax=6, xtick={0.5,1,1.5,3,5}, xticklabels={0.5,1,1.5,3,5},
        ymin=2.0, ymax=2.6, axis y line*=left,
        legend to name=taulegend, legend columns=3,
        legend style={font=\scriptsize, draw=none, column sep=8pt}]
      \addplot[mark=triangle*, color=blue!70!black, thin]
      coordinates {(0.5,2.25) (1,2.43) (1.5,2.47) (3,2.47) (5,2.47)};
      \addlegendentry{\sysname\ speedup}
      \addlegendimage{mark=o, color=red!80!black, thin}
      \addlegendentry{\sysname\ groups per round (right axis)}
      \addlegendimage{mark=diamond, color=gray!80!black, dashed, thin}
      \addlegendentry{Concatenated Block-STM (right panel)}
      \addlegendimage{mark=o, color=black, dotted, thin}
      \addlegendentry{Block-STM (right panel)}
      \addlegendimage{mark=star, color=gray!80!black, thin}
      \addlegendentry{Concatenated graph OCC (right panel)}
      \addlegendimage{mark=square, color=green!60!black, thin}
      \addlegendentry{Graph-aware OCC (right panel)}
    \end{axis}
    \begin{axis}[tauaxis, xmode=log, xmin=0.42, xmax=6, xtick=\empty, ymin=0, ymax=100,
        axis y line*=right, axis x line=none]
      \addplot[mark=o, color=red!80!black, thin]
      coordinates {(0.5,67.8) (1,28.2) (1.5,23.8) (3,22.8) (5,22.7)};
    \end{axis}
  \end{tikzpicture}\hfill%
  \begin{tikzpicture}
    \begin{axis}[tauaxis, xmode=log, xlabel={$\tau_{\mathrm{CV}}$},
        xmin=0.085, xmax=2.4, xtick={0.1,0.3,0.5,1,2}, xticklabels={0.1,0.3,0.5,1,2},
        ymin=2.0, ymax=2.6, axis y line*=left]
      \addplot[mark=triangle*, color=blue!70!black, thin]
      coordinates {(0.1,2.49) (0.3,2.49) (0.5,2.47) (1,2.35) (2,2.12)};
    \end{axis}
    \begin{axis}[tauaxis, xmode=log, xmin=0.085, xmax=2.4, xtick=\empty, ymin=0, ymax=100,
        axis y line*=right, axis x line=none, ylabel={Groups per round}]
      \addplot[mark=o, color=red!80!black, thin]
      coordinates {(0.1,16.3) (0.3,19.5) (0.5,23.8) (1,42.6) (2,97.2)};
    \end{axis}
  \end{tikzpicture}\hfill%
  \begin{tikzpicture}
    \begin{axis}[tauaxis, xlabel={Workers $k$}, ylabel={Speedup},
        xmin=2, xmax=34, xtick={4,8,16,24,32}, ymin=1.0, ymax=2.7]
      \addplot[mark=triangle*, color=blue!70!black, thin]
      coordinates {(4,2.29) (8,2.56) (16,2.21) (24,2.06) (32,1.93)};
      \addplot[mark=diamond, color=gray!80!black, dashed, thin]
      coordinates {(4,2.28) (8,2.59) (16,1.92) (24,1.89) (32,1.69)};
      \addplot[mark=o, color=black, dotted, thin]
      coordinates {(4,1.80) (8,1.98) (16,1.32) (24,1.24) (32,1.22)};
      \addplot[mark=star, color=gray!80!black, thin]
      coordinates {(4,2.08) (8,2.25) (16,1.99) (24,2.06) (32,2.05)};
      \addplot[mark=square, color=green!60!black, thin]
      coordinates {(4,1.70) (8,1.75) (16,1.62) (24,1.57) (32,1.51)};
    \end{axis}
  \end{tikzpicture}\\
  \ref{taulegend}
  \caption{Influences of the hot-resource threshold $\tau_{hot}$ (left), the balance threshold $\tau_{\mathrm{CV}}$
    (middle), and the worker
    count $k$ (right).}
  \label{fig:tau}
\end{figure*}
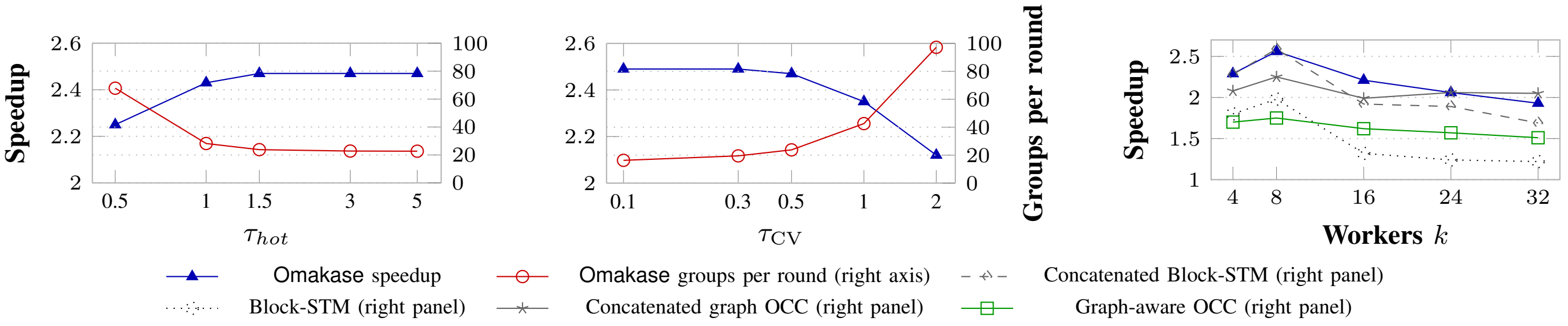

\heading{Intra-block conflict ($\rho_{\mathrm{intra}}$).}
Figure~\ref{fig:artificial} shows that conflict inside a block is what
hurts most. When 10\% of a block's transactions write the hot key, the
two per-block engines reach 5.2$\times$; when 40\% do, only
2.1$\times$, since those transactions must run one after another. Up to
20\% \sysname is no faster: each block alone keeps all eight workers
busy, so the balance test sees no gain from integration. From 30\% on a
single block runs out of independent transactions, \sysname integrates
blocks, and the groups win back most of the loss: 4.25$\times$ against 2.7$\times$ at 30\%, and 3.44$\times$
against 2.1$\times$ at 40\% ($\rho_{\mathrm{inter}}{=}2$).

\heading{Inter-block conflict ($\rho_{\mathrm{inter}}$).}
Conflict between blocks only matters once blocks are integrated, and then it
eats into the gain: at $\rho_{\mathrm{intra}}{=}40\%$, \sysname falls
from 3.44$\times$ when two blocks in a row share hot key $h_1$ to
2.75$\times$ when eight do\edited{: a round of 50 blocks runs as 22 groups at $\rho_{\mathrm{inter}}{=}2$ and 32 at $\rho_{\mathrm{inter}}{=}8$, which reduces the inter-block parallelism available within each group.}

\heading{Why the concatenated round wins here.}
The concatenated round is the fastest engine in Figure~\ref{fig:artificial}. On the artificial
workload the hints are exact, so concatenated graph OCC never
re-executes a transaction, and one window of 5{,}000 transactions offers
more parallelism than \sysname's 20$\sim$30 integrate groups per round. However, we still do not
consider it the better design: \S\ref{sec:speedup} showed that with the
imprecise hints of real Ethereum, conflicting blocks congest a large
window with re-executions (0.88 per transaction) and its slowest round
falls to 1.51$\times$ against 1.86$\times$ for \sysname. Also, even in Figure~\ref{fig:artificial}, 
its advantage gets smaller as conflict grows, to 2.93$\times$ and 2.99$\times$ against
2.75$\times$ at $(\rho_{\mathrm{inter}},\rho_{\mathrm{intra}}){=}(8,40\%)$. 
As we can observe, \sysname is more resilient against the increasing inter-block conflict.

\heading{Hot-resource threshold ($\tau_{hot}$).}
Figure~\ref{fig:tau} measures the influence of both thresholds in \sysname. 
For the hot-resource threshold, speedup rises from 2.25$\times$ at
$\tau_{hot}{=}0.5$ to 2.43$\times$ at $\tau_{hot}{=}1.0$, and then stays flat up to $\tau_{hot}{=}5.0$
(2.47$\times$). A low threshold labels too many keys as hot, so the
integrator refuses integrations that would have improved parallelism and leaves a
100-block round in 68 groups when $\tau_{hot}{=}0.5$ instead of 24 when $\tau_{hot}{=}1.5$. We then set $\tau_{hot}{=}1.5$ by default.

\heading{Balance threshold ($\tau_{\mathrm{CV}}$).}
The integrator stops integrating once a group's simulated load is balanced, i.e.,
with a coefficient of variation below $\tau_{\mathrm{CV}}$ (\S\ref{ss:balance}). A low threshold
keeps integrating up to the cap, which we set to 16 groups per round at 0.1, for 2.49$\times$ speedup. A high threshold, however, declares groups
balanced too early, with 97 groups per round at $\tau_{\mathrm{CV}}{=}2.0$, so speedup falls to 2.12$\times$, close to per-block execution. 
We then set $\tau_{\mathrm{CV}}{=}0.5$ by default, which gives 2.47$\times$.

\heading{Workers ($k$).}
Figure~\ref{fig:tau} (right) varies the worker thread count. Every engine is fastest with 8 workers on this machine, and all
of them slow down when there are more than 8 workers. The two Block-STM-based engines fall faster than the graph-aware OCC engines. 
With more workers, more conflicting transactions execute optimistically and concurrently, but are then
re-executed for validation failure. For example, the concatenated round re-executes transactions 1.60 times per transaction at $k=32$
workers against 0.90 at $k=8$. The three graph-aware OCC engines fall more
gently because the graph keeps transactions from
starting early. What remains is the cost of coordinating more workers, which still makes the engines slower.

\subsection{Why Omakase Works: Fewer Aborts}
\label{sec:why}
\label{sec:aborts}

\begin{table}[t]
  \centering\scriptsize\setlength{\tabcolsep}{2pt}
  \caption{Abort accounting per transaction on real Ethereum workloads. \emph{Cascade} = validation
    failure caused by a lower-indexed transaction's \emph{re-execution}. \emph{From another
    block}: validation failures whose invalidating writer is in a different
    block of the same group.}
  \label{tab:aborts}
  \begin{tabular}{l r rr rr}
    \toprule
    & & \multicolumn{2}{c}{validation failures} & \multicolumn{2}{c}{from another block} \\
    \cmidrule(lr){3-4}\cmidrule(lr){5-6}
    Engine & re-exec & all & cascade & all & cascade \\
    \midrule
    Block-STM              & 0.545 & 0.183 & 0.135 & -- & -- \\
    Concatenated Block-STM & 0.877 & 0.208 & 0.193 & -- & -- \\
    Concatenated graph OCC & 0.545 & 0.294 & 0.224 & 0.088 & 0.073 \\
    Graph-aware OCC        & \textbf{0.268} & \textbf{0.123} & \textbf{0.080} & 0 & 0 \\
    \sysname               & 0.309 & 0.151 & 0.101 & 0.082 & 0.056 \\
    \bottomrule
  \end{tabular}
\end{table}

This subsection answers the third question: why \sysname works, in
terms of aborts. Table~\ref{tab:aborts} counts, per transaction, how
often each engine re-executes a transaction and why, on real Ethereum
workloads.

\heading{Two kinds of abort.}
\edited{A transaction is re-executed either because it failed validation or because, during execution, it read the placeholder of a transaction still being re-executed, or a nonce not yet in place, and gave up at once. Table~\ref{tab:aborts} counts every execution after the first as a re-execution and only the first kind as a validation failure.}
In optimistic execution a transaction fails validation when it read a
value that an earlier transaction later changed. This happens in two
ways. In an \emph{ordinary abort} the earlier transaction simply had not run when the later transaction read.
In a \emph{cascade abort}, the earlier transaction
had already run and was then re-executed. A re-execution creates a new
version of every key it writes, and validation compares versions, not
values, so a transaction that was placed correctly after the first run
still fails even if the value did not change. Cascades are what the argument of \S\ref{sec:overview} is about: one
re-execution can invalidate the transactions that depend on it, and so on.

\heading{Cascades are the common case.}
As shown in Table~\ref{tab:aborts}, three quarters of Block-STM's validation failures are cascades, and for
the concatenated round with block-STM the share is 93\%, because putting a whole round
in one optimistic window turns conflicts between blocks into a concurrent execution that can cascade.

\heading{The conflict graph removes most of them.}
A graph-aware engine does not start a transaction until its RAW parents have executed. 
Running block by block, graph-aware OCC has 41\%
fewer cascades and 51\% fewer re-executions than Block-STM. \sysname
puts several blocks in one group, which brings their conflicts between
blocks back into play, yet it still has 25\% fewer cascades and 43\%
fewer re-executions than Block-STM.

\heading{What is left.}
More than half of \sysname's remaining validation failures come from a
writer in \emph{another block} of the same group (0.082 of 0.151 per
transaction, Table~\ref{tab:aborts}), and two thirds of those (0.056)
are cascades: the reader was placed after the writer's first run, as it
should be, but the writer was later re-executed and its new version
invalidated the read. \edited{The graph orders first executions; it does not prevent a later re-execution from invalidating dependent readers. Group boundaries limit how far these cascades can propagate during concurrent execution.}

\subsection{Stage Costs and Validator Pipeline Throughput}
\label{sec:e2e}

\begin{table}[t]
  \centering\scriptsize\setlength{\tabcolsep}{3pt}
  \caption{Cost of every stage per block (ms), as a fraction of
    sequential execution, and the resulting speedups. }
  \label{tab:e2e}
  \begin{tabular}{l rr rr}
    \toprule
    & \multicolumn{2}{c}{Real Ethereum} & \multicolumn{2}{c}{Synthetic} \\
    \cmidrule(lr){2-3}\cmidrule(lr){4-5}
    Stage & ms/block & of seq. & ms/block & of seq. \\
    \midrule
    Sequential execution (reference) & 9.15 & 1.00 & 13.16 & 1.00 \\
    \midrule
    Pre-execute (proposer)           & 9.32 & 1.02 & 13.38 & 1.02 \\
    Build graph + intra-block reorder (proposer) & 2.02 & 0.22 & 1.57 & 0.12 \\
    Integrate (validator)            & 2.91 & 0.32 & 1.18 & 0.09 \\
    Execute, \sysname (validator)    & 3.71 & 0.41 & 5.76 & 0.44 \\
    \midrule
    Execution only (Table~\ref{tab:absolute}) & 3.71 & \textbf{2.47$\times$} & 5.76 & \textbf{2.28$\times$} \\
    Validator: integrate + execute  & 6.62 & \textbf{1.38$\times$} & 6.94 & \textbf{1.89$\times$} \\
    Every stage on one node, $N{=}100$ proposers & 6.73 & \textbf{1.36$\times$} & 7.09 & \textbf{1.85$\times$} \\
    Block-STM, no preparatory stage  & 4.83 & 1.89$\times$ & 6.53 & 2.02$\times$ \\
    \bottomrule
  \end{tabular}
\end{table}

\heading{Stage costs.}
\edited{This subsection answers the fourth question: what each stage of \sysname costs, and how much of that cost a validator's pipeline can hide by overlapping integration with execution.}
Table~\ref{tab:e2e} gives the cost of each stage per block.
\emph{Pre-execution} costs exactly one sequential run of the block. Every engine that uses hints pays it, but each proposer pays for only its own block, so with $N$ proposers a
validator's share is $1/N$ of it. Building the graph and reordering the transactions add another fifth of a
sequential run. \emph{Integration} is the one stage that only \sysname and concatenated round with graph-OCC have, and every validator pays it. With state in memory it is expensive, taking
2.91\,ms per block on real Ethereum workloads. A validator that
integrates and then executes gets 1.38$\times$ speedup only, below Block-STM's
1.89$\times$. Adding $1/N$ of the proposer stages gives the ``every
stage on one node'' row: with one block per proposer and $N{=}100$, a
validator pays 6.73\,ms per block in all, a speedup of 1.36$\times$ on
real Ethereum and 1.85$\times$ on the synthetic workload.

\begin{table}[t]
  \centering\scriptsize\setlength{\tabcolsep}{4pt}
  \caption{Validator-side cost per block and speedup with the integration
    of round $r{+}1$ on one spare thread while round $r$ executes.}
  \label{tab:overlap}
  \begin{tabular}{l rr}
    \toprule
    & ms/block & speedup \\
    \midrule
    Block-STM, block by block                      & 5.03 & 1.91$\times$ \\
    \sysname, execution only                       & 4.09 & 2.34$\times$ \\
    \sysname, integration only                     & 3.31 & -- \\
    \sysname, integrate then execute               & 7.40 & 1.30$\times$ \\
    \sysname, integration overlapped with execution & 4.29 & \textbf{2.23$\times$} \\
    \bottomrule
  \end{tabular}
\end{table}

\heading{Hide cost by overlapping integration.}
Those rows charge integration and execution one after the other. A
validator can instead integrate round $r{+}1$ on a spare core while its
workers execute round $r$; Table~\ref{tab:overlap} runs all 150 real
rounds back to back both ways. \edited{Overlapping integration with execution raises the measured validator-side cost from 4.09 to 4.29\,ms per block; the 0.20\,ms increment corresponds to hiding approximately 94\% of the 3.31\,ms standalone integration cost. The resulting validator-side speedup is 2.23$\times$, compared with 1.91$\times$ for Block-STM in the same run. These measurements cover validator-side preparation and execution; they do not measure transaction latency through proposal, dissemination, and consensus.}

\section{Conclusion}

Multi-proposer consensus commits a batch of blocks every round, but
today's parallel execution engines still run those blocks one at a
time, so most of the parallelism in a round, and most of a validator's
cores, are idle. This paper presented \sysname, which treats the batch
as the unit of execution. Each proposer pre-executes only its own block
and ships a conflict graph with it, so the cost of building conflict graphs is
spread across all proposers. \edited{At the validator, a greedy integrator considers a block for addition to a group only when its hot-resource set is disjoint from that of the current integrated graph, and accepts the addition only if the balance test passes. A graph-aware OCC engine executes each integrated group, admitting a transaction only after its predicted RAW parents have executed. This reduces aborts caused by optimistic execution. Executing groups sequentially confines cross-block cascades to the current group, although re-executions can still trigger cascades within it.}

On 15{,}000 Ethereum mainnet blocks, about 2.4\,M transactions,
\sysname executes a round 2.47$\times$ faster than sequential execution,
against 1.89$\times$ for Block-STM and 1.66$\times$ for the same engine
run block by block. The concatenated round, which treats the whole round as one
optimistic window is about as fast on average (2.40$\times$) but far
less stable. Its slowest round runs at 1.51$\times$ against 1.86$\times$
for \sysname, and on the five rounds that it runs the slowest, \sysname is 1.4
to 2 times faster. The integration step that \sysname adds costs more
than it saves if a validator waits for it, \edited{but overlapping integration with execution across rounds hides 94\% of the measured integration cost in the validator pipeline.}

In summary, \sysname exploits inter-block parallelism to achieve higher
throughput than existing engines, and keeps stable performance when the inter-block conflicts are high.
With both properties above, \sysname achieves congestion control in transaction
execution for multi-proposer consensus.

\section*{Acknowledgments}
\heading{Generative AI disclosure.} In line with the IEEE policy on
AI-generated content, we disclose that Anthropic's Claude (via Claude
Code) was used under the authors' direction for the text, the figures and
images, and the code of this work. No measurement, result or citation was
generated by AI. The authors reviewed and verified every change and take
full responsibility for the content.

\bibliographystyle{IEEEtran}
\bibliography{references}

\end{document}